\UseRawInputEncoding
\documentclass[10pt,aps,prd,reprint,nofootinbib,floats,floatfix,amsfonts,amssymb,amsmath,preprintnumbers,notitlepage,superscriptaddress]{revtex4-1}

\usepackage{graphicx} 
\usepackage[usenames]{xcolor}
\usepackage{hyphenat}
\usepackage{url}
\usepackage[normalem]{ulem}
\usepackage{units}
\usepackage{soul}

\newcommand{\Msun}{$\mathrm{M_{\odot}}$}

\begin{document}

\begin{flushleft}
KCL-PH-TH/2024-24
\end{flushleft}

\title{Rapid detection of gravitational waves from binary black hole mergers using sparse dictionary learning}

\author{Charles Badger$^*$}
\affiliation{Theoretical Particle Physics and Cosmology Group,  Physics Department, \\ King's College London, University of London, Strand, London WC2R 2LS, United Kingdom}
\author{Rahul Srinivasan$^*$}
\affiliation{Universit\'e C\^{o}te d'Azur, Observatoire de la C\^{o}te d'Azur, CNRS, Laboratoire Lagrange, Bd de l'Observatoire, F-06304 Nice, France}
\thanks{Authors contributed equally.}

\affiliation{Universit\'e C\^{o}te d'Azur, Observatoire de la C\^{o}te d'Azur, CNRS, Artemis, Bd de l'Observatoire, F-06304 Nice, France}
\affiliation{SISSA, Via Bonomea 265, 34136 Trieste, Italy and INFN Sezione di Trieste}
\affiliation{IFPU - Institute for Fundamental Physics of the Universe, Via Beirut 2, 34014 Trieste, Italy}

\author{Alejandro Torres-Forn\'e}
 \affiliation{Departamento de Astronom\'ia y Astrof\'isica, 
 Universitat de Val\`encia, Dr.~Moliner 50, 46100 
 Burjassot (Val\`encia), Spain}
\affiliation{Observatori Astron\`omic, Universitat de
 Val\`encia, Catedr\'atico Jos\'e Beltr\'an 2, 
 46980 Paterna (Val\`encia), Spain}%

\author{Marie Anne Bizouard}
 \affiliation{Universit\'e C\^{o}te d'Azur, Observatoire de la C\^{o}te d'Azur, CNRS, Artemis, Bd de l'Observatoire, F-06304 Nice, France}

\author{Jos\'e A.~Font}
 \affiliation{Departamento de Astronom\'ia y Astrof\'isica, 
 Universitat de Val\`encia, Dr.~Moliner 50, 46100 
 Burjassot (Val\`encia), Spain}%
 \affiliation{Observatori Astron\`omic, Universitat de
 Val\`encia, Catedr\'atico Jos\'e Beltr\'an 2, 
 46980 Paterna (Val\`encia), Spain}%

\author{Mairi Sakellariadou}
\affiliation{Theoretical Particle Physics and Cosmology Group,  Physics Department, \\ King's College London, University of London, Strand, London WC2R 2LS, United Kingdom}

\author{Astrid Lamberts}
\affiliation{Universit\'e C\^{o}te d'Azur, Observatoire de la C\^{o}te d'Azur, CNRS, Laboratoire Lagrange, Bd de l'Observatoire, F-06304 Nice, France}
\affiliation{Universit\'e C\^{o}te d'Azur, Observatoire de la C\^{o}te d'Azur, CNRS, Artemis, Bd de l'Observatoire, F-06304 Nice, France}

\date{\today}

\begin{abstract}
    Current gravitational wave (GW) detection pipelines for compact binary coalescence based on matched-filtering have reported over 90 confident detections during the first three observing runs of the LIGO-Virgo-KAGRA (LVK) detector network. Decreasing the latency of detection, in particular for future detectors anticipated to have high detection rates, remains an ongoing effort. In this paper, we develop and test a sparse dictionary learning (SDL) algorithm for the rapid detection of GWs. We evaluate the algorithm’s biases and estimate its GW detection rate for an astrophysical population of binary black holes. The SDL algorithm is assessed using both, simulated data injected into the proposed A+ detector sensitivity and real data containing confident detections from the third LVK observing run. We find that our SDL algorithm can reconstruct a single binary black hole signal in less than 1\,s. This suggests that SDL could be regarded as a promising approach for rapid, efficient GW detection in future observing runs of ground-based detectors.
\end{abstract}

\maketitle

\section{Introduction}
\label{sec:intro} 

The 2015 detection of gravitational wave (GW) source GW150914 marked a new era in physics, and signaled the rise of GW 
astrophysics~\cite{FirstGWDetec}. 
Since then, more than 90 GW signals from compact binary coalescences (CBC) have been confidently detected with the Advanced LIGO~\cite{Aasi:2014jea} and Advanced Virgo~\cite{Acernese:2014hva} ground-based detector network~\cite{O1O2Search_GWOSC, O3Search_GWOSC}, recently having added KAGRA~\cite{KAGRA:2020tym}. Detections are not only essential in understanding the properties of individual CBCs, but also their underlying populations in our local Universe. In the third observing run, it was found that the binary black hole (BBH) mass distribution has localized over- and under-densities relative to a power-law distribution with peaks emerging at chirp masses of $8.3^{+0.3}_{-0.5}~M_{\odot}$ and $27.9^{+1.9}_{-1.8}~M_{\odot}$, with a merger rate proportional to $(1+z)^{\kappa}$ where $\kappa = 2.9^{+1.7}_{-1.8}$ for redshift $z < 1$~\cite{GWTC-3_Population_Properties}.

A number of detection pipelines have been developed to statistically determine the likelihood of the presence of a GW signal in detector data - most notably the 
GstLAL~\cite{sachdev2019gstlal}, 
MBTA~\cite{Aubin_2021}, 
PyCBC~\cite{Dal_Canton_2021}, 
IAS~\cite{Venumadhav:2019lyq}, 
SPIIR~\cite{chu2021spiir}, and 
cWB~\cite{Klimenko_2016} pipelines. Most CBC pipelines are based on matched-filtering in which incoming GWs are cross-correlated to
pre-computed signal templates (or waveform approximants) for given
source parameters (see e.g.~\cite{LIGOScientific:2019hgc}). Despite their many strengths, development of faster, more computationally efficient procedures to search for GW signals in data grows increasingly important as data volume increases. 
More recently, machine learning methods and artificial intelligence techniques have been developed to improve GW detection prospects (see e.g.~\cite{Kim_2015,George_2018,Marx:2024} for specific proposals and~\cite{Cuoco_2020, app13179886, Sch_fer_2023,Stergioulas:2024} for reviews on applications of machine learning in GW astronomy).

In this study we apply one such machine learning approach -- sparse dictionary learning (SDL) -- to investigate the viability of this method to detecting CBCs buried in the strain data of ground-based detectors. Our work builds on the study initiated in~\cite{Charlie+2023} where SDL was first employed to detect signals from massive black hole binary mergers in the presence of foreground noise due to Galactic binaries with LISA. A salient feature of the analysis reported here is that its focus is on the prospects of detecting GW signals from an astrophysical population of BBH mergers simulated from the catalog of mergers produced in~\cite{Srinivasan+23}. For completeness, we also evaluate the intrinsic performance of our SDL algorithm when  considering a population of BBH mergers with a flat prior over a broader range of parameters. We use only single-detector data and the dictionary is trained with simulated waveforms based on the \texttt{IMRPhenomD} approximant from the \texttt{PyCBC} library~\cite{alex_nitz_2024_10473621}. The performance of our SDL algorithm is evaluated by injecting the data into the simulated sensitivity of the proposed A+ detector~\cite{KAGRA:2013rdx} and also using real data
containing confident detections from the third LVK observing run. 

Our study indicates that SDL is an encouraging technique to detect BBH signals buried in detector data. The relatively few training signals needed to create a dictionary and its quick reconstruction speed make SDL a potentially ideal approach for a rapid and computationally efficient detection pipeline.  We are able to reconstruct injected waveforms in simulated A+ noise~\cite{KAGRA:2013rdx} and those corresponding to real O3b confident events with FAR $\leq 12\,\rm{yr}^{-1}$, each within less than 1 s. These results suggest that the development of an SDL-based, full-fledged detection pipeline for ground-based GW interferometers is worth pursuing. 

The rest of the paper is organised as follows. In Sec.~\ref{sec:methods} we describe the methodology used to access the dictionary approach where in Sec.~\ref{sec: DL_Approach} we first introduce the dictionary learning approach to be applied to astrophysical populations described in Sec.~\ref{sec: test_pop}. We present our results in Sec.~\ref{sec:results}. 
Lastly, we discuss further prospects of the dictionary learning approach and summarize our findings in Sec.~\ref{sec:discussion}.

\section{Methodology}
\label{sec:methods}

\subsection{Sparse Dictionary Learning}
\label{sec: DL_Approach}

The development of algorithms for the sparse reconstruction of a signal over a dictionary has received significant interest in the last decades~\citep{Chen:2001,Elad:2006,Mairal:2012}.  This approach is an alternative to more traditional signal representations based on Fourier decomposition, wavelets, chirplets, or warplets. The use of SDL for GW data analysis was first introduced in~\cite{Alex+2016_GWDenoising}. It has since been applied in a number of subsequent works in the field of GW astronomy~\cite{Miquel:2019,Alex+2020,Saiz-Perez2022,Charlie+2023,Powell2023}. Following~\cite{Alex+2016_GWDenoising} we model the detector strain, $s(t)$, as a superposition of the CBC signal $h(t)$ and the detector noise $n(t)$:
\begin{equation}
    s(t)=h(t)+n(t).
\end{equation}
The objective of SDL~\cite{Mallat:1993,dict_learning} is to find a sparse vector $\alpha$ that reconstructs the true signal $h$ as a linear combination of columns of a preset matrix $\textbf{D}$, called dictionary,
\begin{equation}
h \sim \textbf{D} \alpha.
\end{equation}
The columns of the dictionary, called atoms, can be a set of prototype signals, like Fourier basis or wavelets, or one can design the dictionary to fit a given set of GW templates. In our study, those signals are BBH waveforms.

The loss function is expressed as
\begin{equation}
\label{eq:constrain}
    J(h) = ||s-h||_{L_2}^2 + \lambda \mathcal{R}(h),
\end{equation}
and searches for a solution that minimises $J(h)$ in the time domain, where $||\cdot||_{L_2}$ is the $L_2$ norm~\cite{10.1137/07070156X,Mairal:2009}.
The first term in the loss function, often referred to as the {\sl error term}, measures how well the solution fits the data, while the regularisation term $\mathcal{R}(h)$ captures any imposed constraints. 
The regularisation parameter 
$\lambda$ tunes the weight of the regularisation term relative to the error term; it is a hyperparameter of the optimisation process. 

We apply a learning process where the dictionary is trained to fit a given set of signals. For our CBC signals the waveforms are aligned at the strain maximum (i.e.~at the time of merger for each signal) and divided into patches, with the number of patches ($p$) much larger than the length of each patch ($w$). To train the dictionary we consider both the sparse vector $\alpha$ and the dictionary $\textbf{D}$ as variables:
\begin{equation}
\label{eq:dict_learning}
\alpha_{\lambda}, \textbf{D}_{\lambda}=\underset{\alpha, \textbf{D}}{\rm{argmin}} \left\{\frac{1}{w}\sum_{i=1}^{p}||\textbf{D}\alpha_i- {x}_i||^2_{L_2}+\lambda ||\alpha_i||_{L_1}\right\},
\end{equation}
with $x_i$ denoting the $i$-th training patch and vector sparsity imposed via the regularisation term $\mathcal{R}(h)=||\alpha||_{L_1}$, using the $L_1$ norm. We note that $\alpha_{\lambda}, \textbf{D}_{\lambda}$ cannot be solved simultaneously unless the variables are considered separately as outlined in~\cite{Mairal:2009}. This is commonly called the ``basis pursuit''~\cite{BasisPursuit_Chen+2001, basis_pursuit}  or ``least absolute shrinkage and selection operator''
(LASSO)~\cite{Lasso_Tibshirani1996} problem. Once we have trained dictionary $\textbf{D}_{\lambda}$, we find sparse vector $\alpha_{\rm min}$ that minimises Eq.~(\ref{eq:constrain}) to retrieve reconstructed waveform $h_r = \textbf{D}_{\lambda}\alpha_{\rm min}$.

In our study we will be surveying a range of BBH signals with optimal signal-to-noise ratio (SNR) $\rho_{\rm{opt}}$ defined as
\begin{equation}
    \rho_{\rm{opt}} = \sqrt{(h|h)},
\end{equation}
for a deterministic waveform $h$ where 
\begin{equation}
    (x|y)=2 \int^\infty_0 \frac{x(f) y^*(f) + x^*(f) y(f)}{S_n(f)} {\rm{d}}f\,,
\end{equation}
and where 
$S_n(f)$ is the one-sided noise power spectral density (PSD) and symbol $\ast$ denotes complex conjugation.
To measure the performance of a dictionary, we calculate the overlap between detector strain $s$ and the recovered waveform $h_r$,
\begin{equation}
    \mathcal{O}(s, h_r) = \frac{(s|h_r)}{\sqrt{(s|s)(h_r|h_r)}}.
\end{equation}
The overlap $\mathcal{O}$ can range between -1 and 1, with 1 reflecting perfectly matched signals, and -1 implying perfect anti-correlation. The overlap is widely used in the  GW community for identifying transient CBC signals through matched filtering using waveform template banks \cite{Cutler:1994ys,Cornish:2014kda, LIGOScientific:2019hgc, Cornish:2020dwh}. This calculation yields information on the quality of the signal reconstruction in the presence of detector noise. 

\subsection{BBH populations}
\label{sec: test_pop}

The SDL algorithm is employed to reconstruct an astrophysical population of BBH mergers. To construct this population we use a year of observation based on the catalog of mergers described in the \textit{default} model of \cite{Srinivasan+23}. The black holes have source frame masses ranging from 5 \Msun\ to 45 \Msun \ from stars with metallicity between 1\%-100\% of the Solar metallicity, and a formation redshift between 0 to 8. 
Compared to other merger catalogs, this one is built on a comprehensive 3-dimensional model of the binary star formation rate in the Universe as a function of progenitor galaxy properties (mass and metallicity) and the redshift of formation. The efficiency in forming merging black holes is derived using the \textit{default} prescription of the rapid binary evolution population synthesis code COSMIC (v3.4.0) \citep{Breivik:2019lmt}. The pair-instability supernova mechanism in massive stars enforces the upper mass limit. Moreover, BBHs formed from isolated binary evolution tend to favor equal mass ratio systems.
We note that the local astrophysical merger rate of the catalog is higher than that quoted by the LVK Collaboration (see Fig.~5 in~\cite{Srinivasan+23}).

We investigate our detection biases by evaluating the performance of the SDL algorithm with a flat-prior testing population of 150,000 BBH signals. Compared to the astrophysical population (and also the {\it training} population we use to optimize our dictionary hyperparameters; see below), the flat-prior explores a larger range in the black hole parameters: total source frame mass $5 ~\rm{M_{\odot}} \leq \rm{M_{tot}} \leq 80 ~\rm{M_{\odot}}$, redshift $0 \leq z \leq 8$, and mass ratio $1 \leq q \leq 5$. 
Establishing the method's robustness in detecting systems outside the training dataset validates that it does not overfit. Moreover,  building a flat-prior population demonstrates the algorithm's performance in detecting GWs from BBHs outside the considered astrophysical population.

\subsection{Dictionary optimisation}
\label{sec: dict_opt}

Optimal signal reconstruction quality can only be achieved after the hyperparameters of the dictionary have been determined, namely the patch length $w$, the number of patches $p$, and the regularisation parameter $\lambda$. As done in previous works~\cite{Alex+2016_GWDenoising}, we define a suitable set of hyperparameters as one that yields the best results according to a given quality metric. Therefore, we manually check a large range of dictionaries, using the signal overlap as our quality metric, to find the best dictionary in this set for our purposes.  

We first build a dictionary that attempts to reconstruct an underlying CBC signal in detector data using \textit{noise-free} training signals. There is no analytical method to determine the number of training waveforms needed to sufficiently train a dictionary, but one must include a minimal number to cover a desired parameter space efficiently. To this aim we generate three separate sets of 100, 150, and 200 training BBH waveforms (or atoms), each set with uniformly distributed primary masses from 3 \Msun\ to 60 \Msun, secondary source frame masses selected such that almost all mass ratios are close to $1$, uniform sky location, coalescence phase, and uniformly distributed redshift from 0 to 8 to formulate our dictionary. For simplicity, we consider zero spin black holes. All waveforms capture the merger, inspiral and ringdown of the signal projected in the LIGO Hanford detector lasting 0.375~s at sampling frequency $f_s = 8192$ Hz. This is a highly conservative sampling frequency as most BBH signal power is concentrated at lower frequencies - computational efficiency could be improved with downsampling. They are generated using the waveform approximant \texttt{IMRPhenomD} as implemented in the \texttt{PyCBC} 
library~\cite{alex_nitz_2024_10473621}. These are then whitened using detector noise sensitivity and are then trained using Eq.~(\ref{eq:dict_learning}) to create a learned dictionary. 

\begin{figure}[t]
    \centering
    \includegraphics[width=\linewidth]{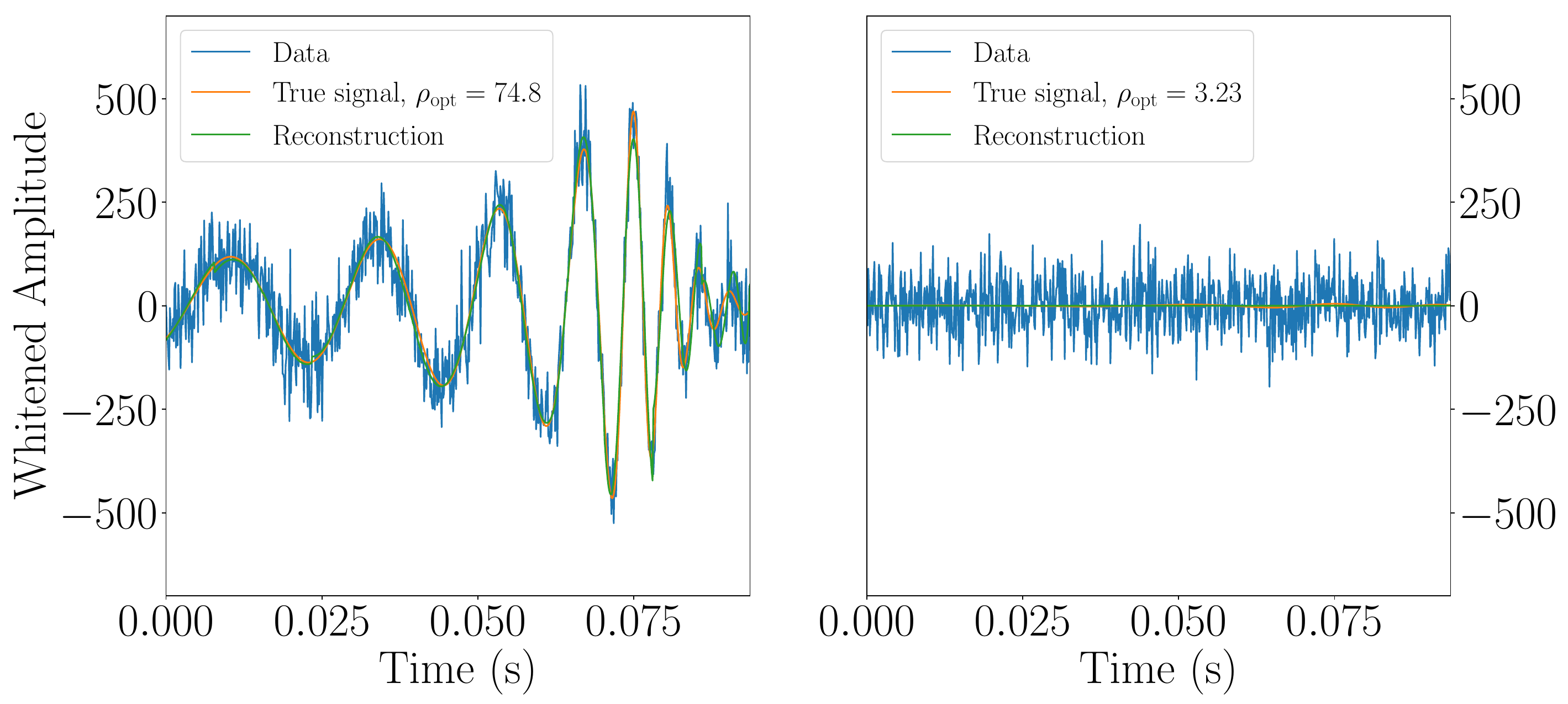}
    \caption{
    Examples of signal reconstruction (green curve) using a trained dictionary for a given detector data (blue curve). The true GW signal (orange curve) present in the detector data is shown for comparison. In the two panels, we illustrate the reconstructions for different optimal SNRs $\rho_\mathrm{opt}$ (shown in legend). The stronger example is reconstructed with an overlap of 0.57 and the weaker one with an overlap of 0.21.}
\label{fig:DictionaryReconstructionExamples}
\end{figure}

We build dictionaries of patch length $w = [2, ~2^9]$ in powers of 2 and number of patches $p = [w, ~10w]$ in $0.5w$ intervals. This leads to a total of 513 dictionaries. Each of them takes between 5 s to 454~s to be created \footnote{Created using Intel E5-2698 v4 CPU model.}, increasing in time as patch length increases. 
Once a dictionary has been built, one must determine its performance quality using validation signals. We generate an additional 25 CBC waveforms injected into detector noise over the same parameter space. Finally, we reconstruct an additional set of 25 CBC testing signals over $\lambda = [10^{-6}, ~10^{-1}]$, as done in previous works on SDL~\cite{Alex+2016_GWDenoising,Miquel:2019,Alex+2020,Saiz-Perez2022,Charlie+2023,Powell2023}. The calibration of hyperparameter $\lambda$ is particularly important as too large a value would result in a failure to reconstruct (returning mainly zeros) whereas too small a value would leave the input data unaltered. We use the overlap to determine the quality of reconstructed signals for different dictionaries, finding that the combination $w=512$, $p=1.5w=768$, $\lambda = 10^{-4}$ for dictionaries with 150 and 200 training signals gives the largest overlap. With this, we fix these hyperparameters for the 150 training signal dictionary\footnote{This was selected for smaller efficiency, using smaller storage space compared to a 200 training signal dictionary.} for the rest of the study. Examples of BBH reconstructions are shown in Fig.~\ref{fig:DictionaryReconstructionExamples}. One can see that when the injected BBH waveform is sufficiently strong (left) the amplitude and the frequency content are well reconstructed. However, for the notably weaker signal shown on the right panel the waveform reconstruction closely approximates a flat line.

\subsection{SDL detection pipeline}
\label{sec: det_meth} 

\begin{figure}
    \centering
    \includegraphics[width=1.\linewidth]{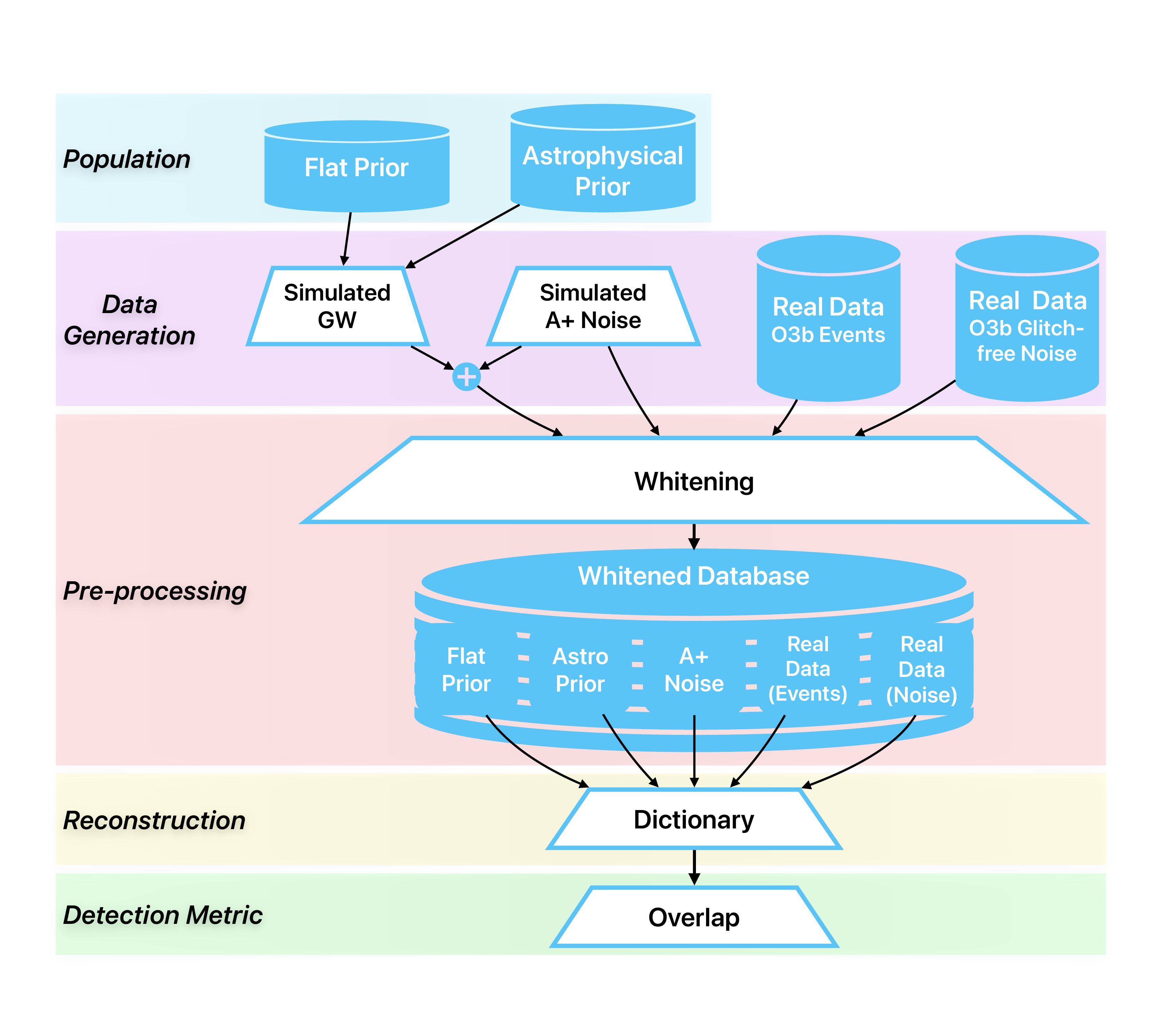}
    \caption{ 
    Flowchart illustrating the procedures for analyzing the BBH populations (flat and astrophysical) and the real O3b data.}
    \label{fig:ML-flowchart}
\end{figure}

The flowchart plotted in Fig.~\ref{fig:ML-flowchart} summarizes the various steps involved in our SDL detection pipeline.
All testing data (flat prior and astrophysical prior populations) are designed to have the same observation time and sampling rate as the noise-free training signals. 
We whiten the data using the detector PSD, divide it into 0.375~s intervals and reconstruct the strain using a learned dictionary. We then compute the overlap of the noise strain and the reconstruction. The resulting overlap values comprise the single-detector noise distribution. 

To compute the false alarm rate (FAR) of the single-detector analysis, i.e.~the rate that an event is at or above a statistical threshold in data solely consisting of noise, we consider datasets that include detector noise. The FAR is estimated as the fraction of noise-only data windows identified as CBC signal per unit of time using a long time period of detector noise. The FAR is limited by the observation period $T_\mathrm{obs}$, being at best $1/T_\mathrm{obs}$. For this study, we consider one month of data. Stationary, Gaussian detector noise can be simulated using the \texttt{PyCBC} module. In addition, in case of real data, we select periods of data such that there are no confidently detected BBHs present in the chosen GPS time windows.

\begin{figure*}
    \centering
    \includegraphics[width=0.45\linewidth]{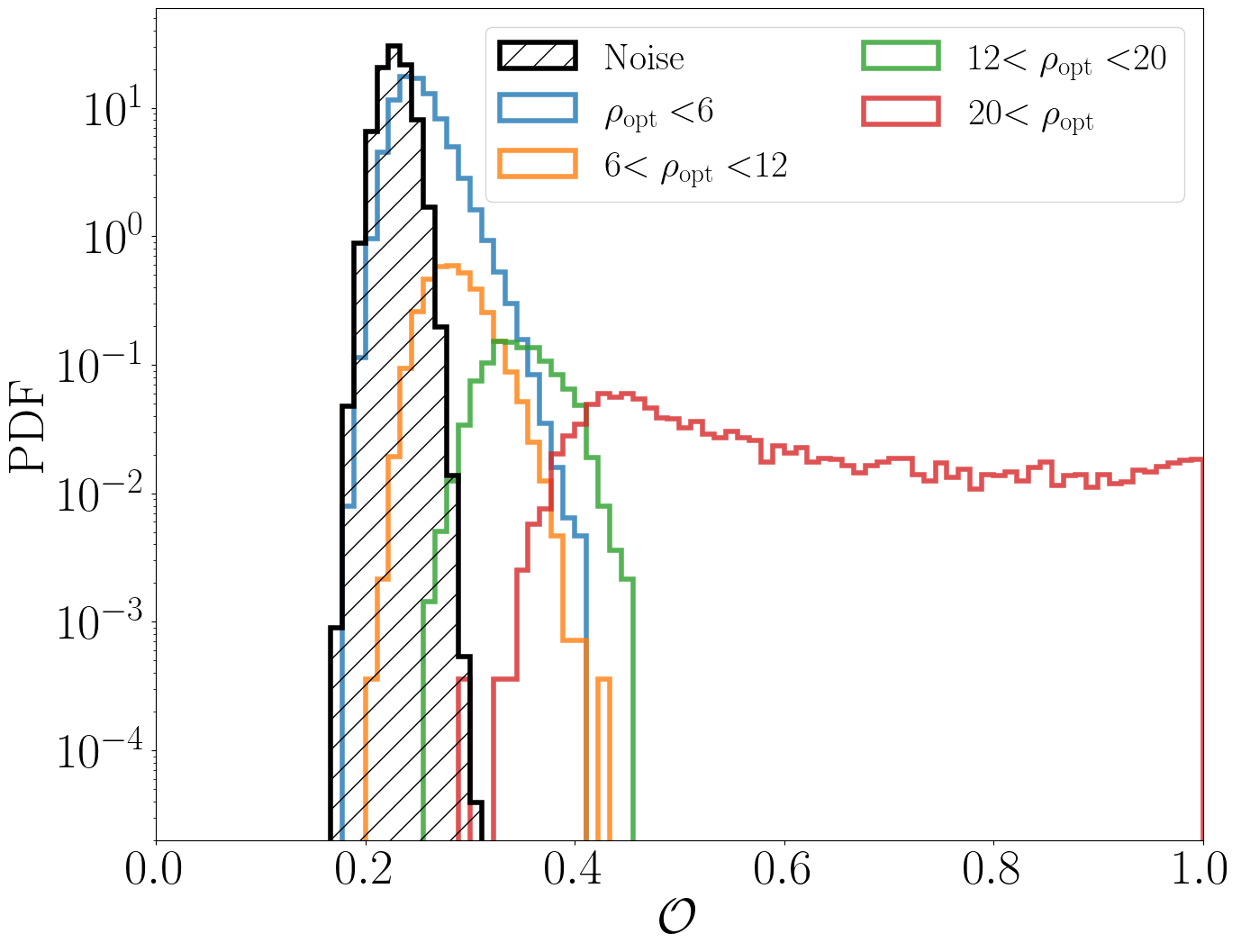}
    \includegraphics[width=0.45\linewidth]{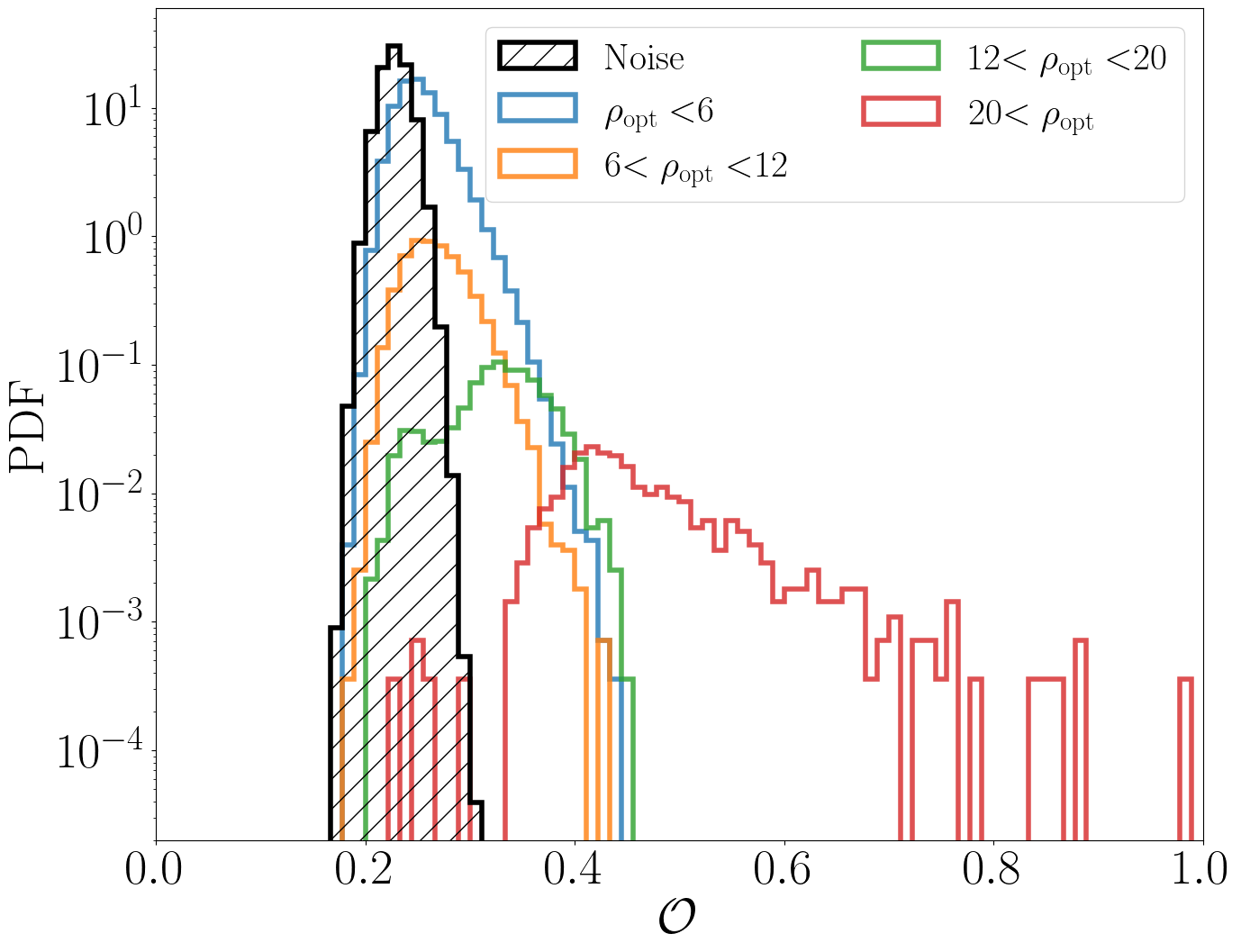}\\
    \caption{ 
    Overlap distributions for the flat-prior (left) and astrophysical (right) populations in comparison to simulated A+ noise (black). The different colors in the distributions correspond to different ranges in the optimal SNR of the injected signals (as indicated in the legends).}
    \label{fig:AplusResults}
\end{figure*}

\section{Results}
\label{sec:results} 

Once a suitable dictionary has been found (see Section~\ref{sec: dict_opt}) we survey two types of datasets. We first apply our SDL pipeline to a suite of BBH waveforms from an astrophysical population~\cite{Srinivasan+23} and from a flat prior population, in both cases using the LIGO Hanford detector with A+ sensitivity, and then apply the method to current O3b LIGO Hanford detections. 

\subsection{BBH detection with simulated A+ sensitivity}
\label{sec: AplusResults}



\begin{figure}[b]
    \includegraphics[width=\linewidth]{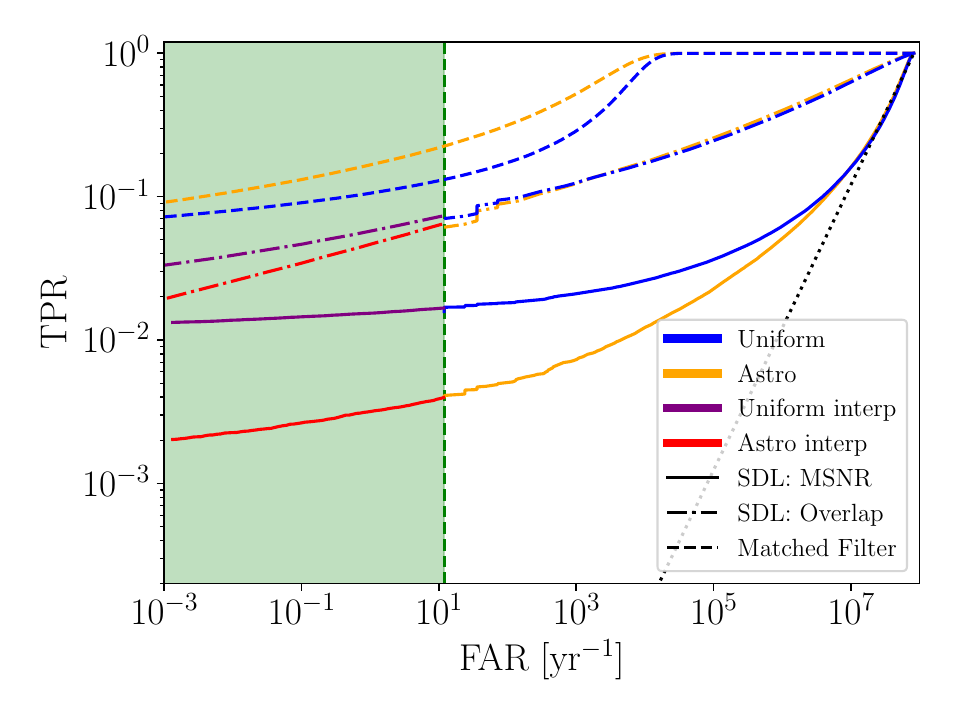}
    \caption{Receiver Operating Characteristics of the SDL pipeline for both detection metrics: overlap in dash-dotted and MSNR in solid lines. The orange (blue), represent the true positive rate (TPR) as a function of the FAR for events in the astrophysical (flat) prior dataset, considering the A+ sensitivity of the Hanford detector. The green background indicates the FAR regime where SDL true positive rate is extrapolated. The dashed lines are the true positive rate of a single-detector matched filter search covering the same CBC parameter space than the SDL search.}
    \label{fig:Roc}
\end{figure}

We first analyze the detection prospects in the LIGO-Hanford detector with upgraded A+ sensitivity. We plot in Fig.~\ref{fig:AplusResults} the overlap distributions for the two BBH populations considered. The various distributions shown correspond to different ranges in the optimal SNR of the injected signals. In addition, the normal distribution shown in black corresponds to the noise-only overlap. For both populations almost all injections with $\rho_{\rm{opt}}>20$ have overlaps beyond the detector noise. Both populations have similar overlap distributions up to $\rho_{\rm{opt}}<20$. However, beyond this SNR, the astrophysical distribution sharply falls, whereas the flat-prior population plateaus. The former is attributed to the dearth of high $\rho_{\rm{opt}}$ astrophysical events due to far fewer BBHs with increasing distance (and redshift). In contrast, the flat-prior distributes BBHs uniformly across distance and, hence, $\rho_{\rm{opt}}$.

Fig.~\ref{fig:Roc} shows the Receiver Operating Curves (ROCs) both for the astrophysical and flat-prior population using the overlap metrics (overlap), as well as the matched SNR (MSNR) of the reconstructed waveform by the SDL pipeline. For FAR lower than $12\,\mathrm{yr^{-1}}$, the SDL ROCs are extrapolated.
The true positive rate is systematically larger for the overlap metric, although for low FAR, the performance tends to be similar ($0.1 - 1$\% for MSNR to be compared to $1-2$\% for overlap at a FAR of $10^{-3}~\mathrm{yr^{-1}}$).
We have found that the MSNR metric underperforms because the SDL method underestimates the waveform's amplitude. The overlap compares only the phases of two waveforms, whereas the matched SNR additionally takes into account the similarities in amplitude. Thus, the underestimated amplitude of the reconstructed waveform weakens the matched SNR statistic in comparison to the overlap. To this end, we choose the overlap as our detection metric in this study and discuss the remainder of our results with this metric.
We also compare SDL detection performances with those of a matched filter (MF) search. We build a single-detector MF search comprising 4582 templates over the same parameter space as the SDL search with a minimal-match of 97\% using a 3.5 post-Newtonian approximant waveform in \texttt{pycbc\_geom\_nonspinbank} function of the \texttt{PyCBC} library~\cite{PhysRevD.60.022002, Cokelaer:2007kx, PhysRevD.87.024033}. The FAR of the template bank search is computed using Eq. B2 in~\cite{Capano_2014}.


The MF search  outperforms the SDL results.
The universe volume comparison of the  SDL and MF methods, approximated as $\langle \rm{SNR}_{\rm{SDL}}^3 \rangle / \langle \rm{SNR}_{\rm{MF}}^3 \rangle$, for the astrophysical data set is found to be $0.81$, and $0.66$ for the flat-prior. This implies that there is a 19\% loss of detectable sources using the SDL method opposed to a MF search for the astrophysical data set. 

\begin{figure}[t]
    \includegraphics[width=\linewidth]{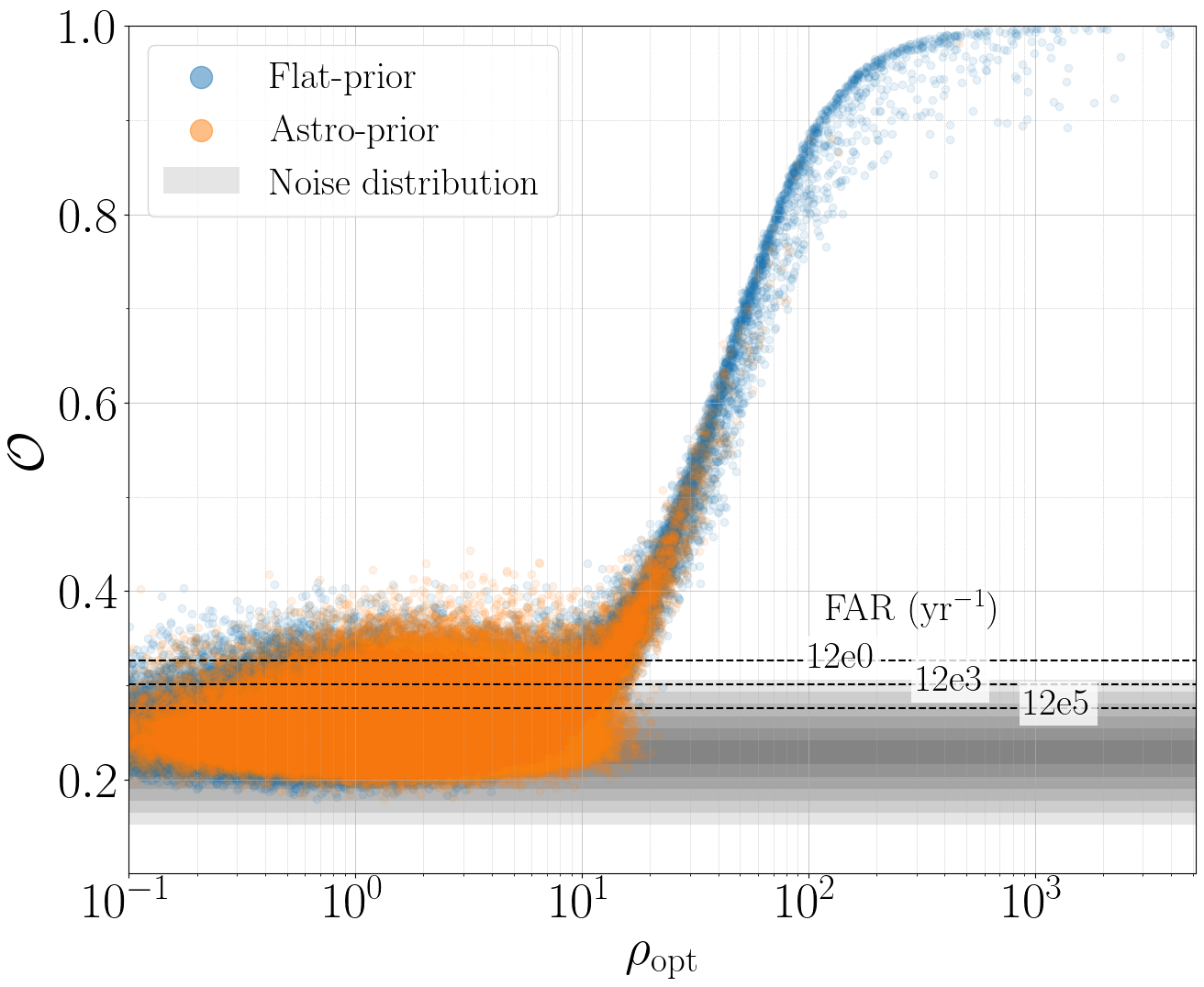}
    \caption{Reconstructed overlap $\mathcal{O}$ for both the astrophysical and flat prior populations as a function of injected SNR $\rho_{\rm{opt}}$. The shaded grey regions in decreasing transparency correspond to the $1\sigma$ to $6\sigma$ regions of the noise distribution obtained with LIGO Hanford simulated data with the A+ sensitivity.}
    \label{fig:Overlap_vs_SNR}
\end{figure}

Fig.~\ref{fig:Overlap_vs_SNR} displays a scatter plot of the overlap for both populations as a function of optimal SNR $\rho_{\rm{opt}}$. One sees that irrespective of the dataset, injections with $\rho_{\rm{opt}} \geq 15$ have an $\rm{FAR} < 12 \,\rm{yr}^{-1}$ (or $\mathcal{O}(s, h_r) \geq 0.35$). This corresponds to a confidence level of more than $5\sigma$ from the noise distribution, as depicted by the shaded grey regions. Therefore, injections in the LIGO Hanford detector with A+ sensitivity and $\rho_{\rm{opt}} \geq 15$  will be confidently detected with our SDL pipeline. As expected, injecting signals of increasingly larger SNR improves reconstruction quality and detection prospects. A near perfect reconstruction $\mathcal{O}(s, h_r) \geq 0.95$ is achieved for injections with $\rho_{\rm{opt}} \geq 100$.

\begin{figure}[t]
    \includegraphics[width=\linewidth]{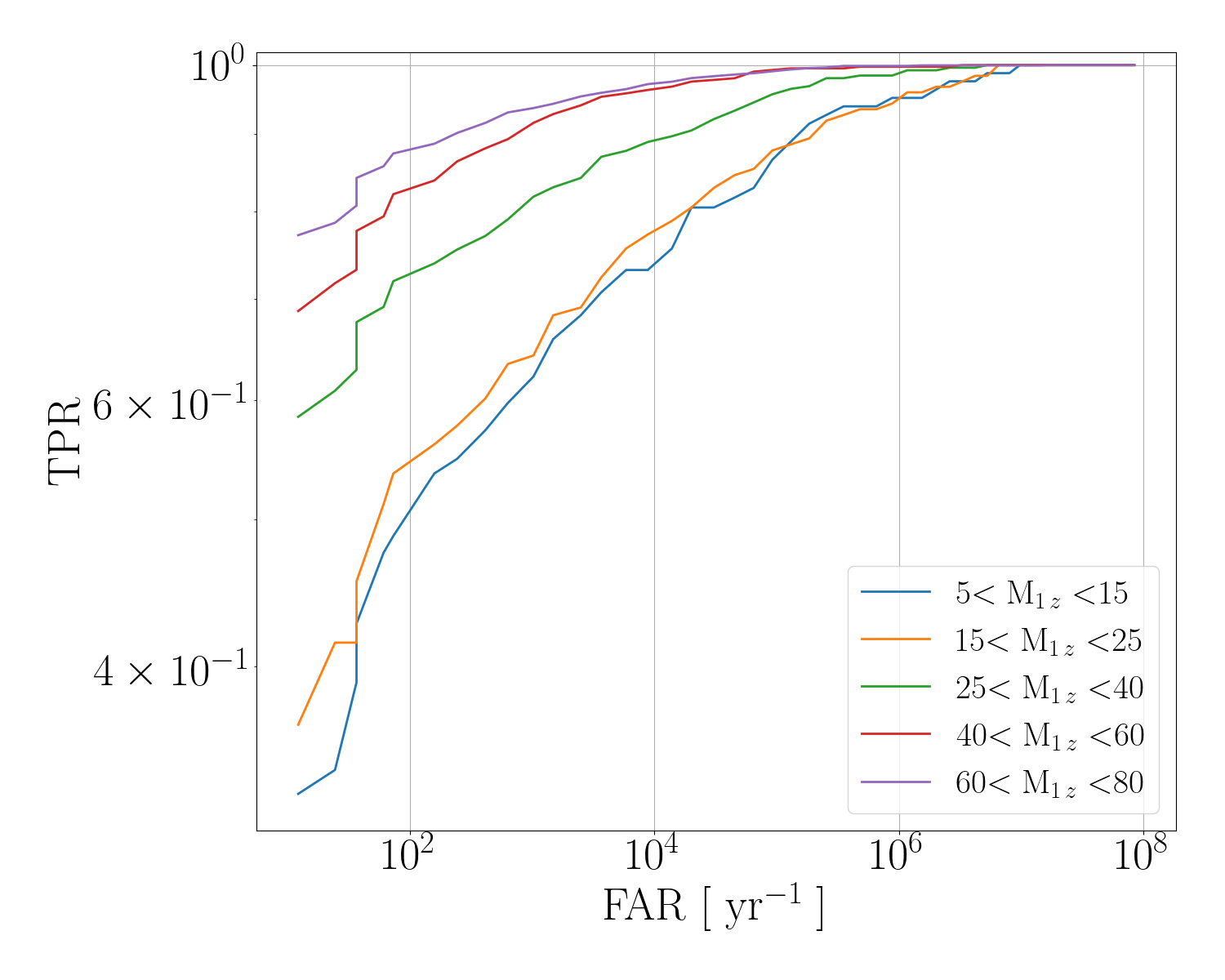}
    \caption{Receiver operating characteristics of the SDL pipeline for BBH systems with an SNR = 12.5$\pm$ 2.5, for different primary black hole redshifted mass ($\mathrm{M}_{1,z}$). The true positive rate is shown as a function of the FAR for events distributed in the flat-prior dataset, added to simulated LIGO Hanford detector noise with A+ sensitivity.}
    \label{fig:Roc_m1}
\end{figure}

ROCs for systems with different primary BBH redshifted masses are depicted in Fig.~\ref{fig:Roc_m1}, to show the true positive rate as a function of the primary BH's detector-frame mass for events with nearly the same SNR (=12 $\pm$ 2.5). Larger masses result in flatter ROC curves. This can be attributed to the frequency evolution as a function of BH mass. As discussed above, the frequency of the waveform plays a role in reconstruction quality  (cf.~Fig.~\ref{fig:DictionaryReconstructionExamples}). Smaller detector-frame mass systems are more difficult to reconstruct accurately due to the higher merger frequency - the predominantly sinusoidal signal within the 0.375~s of input data is difficult for our algorithm to distinguish in comparison to a chirping signal of higher mass systems. This suggests that our choice of dictionary caters more to large mass BBH systems.

\begin{figure}
    \includegraphics[width=\linewidth]{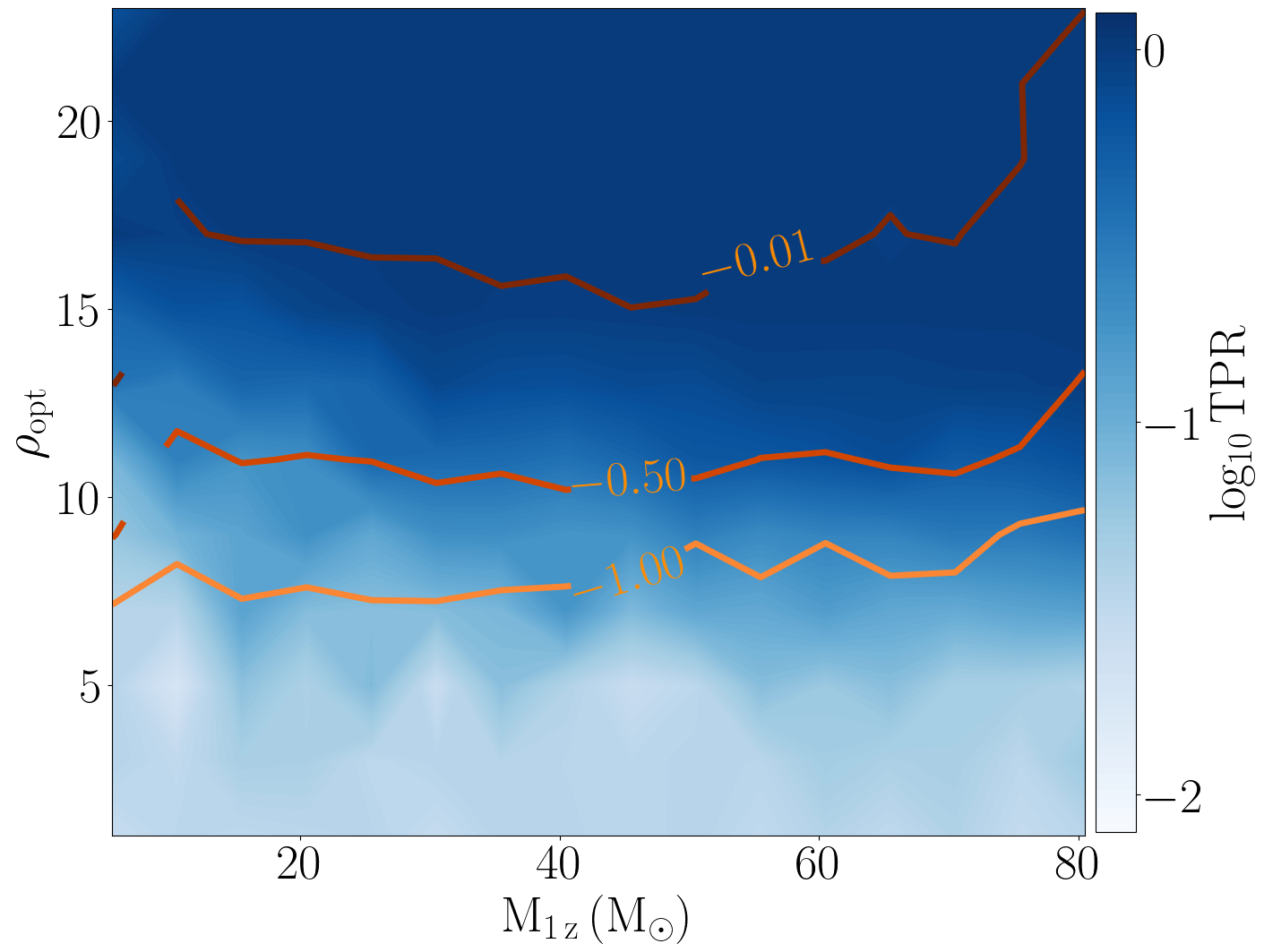}\\
    \includegraphics[width=\linewidth]{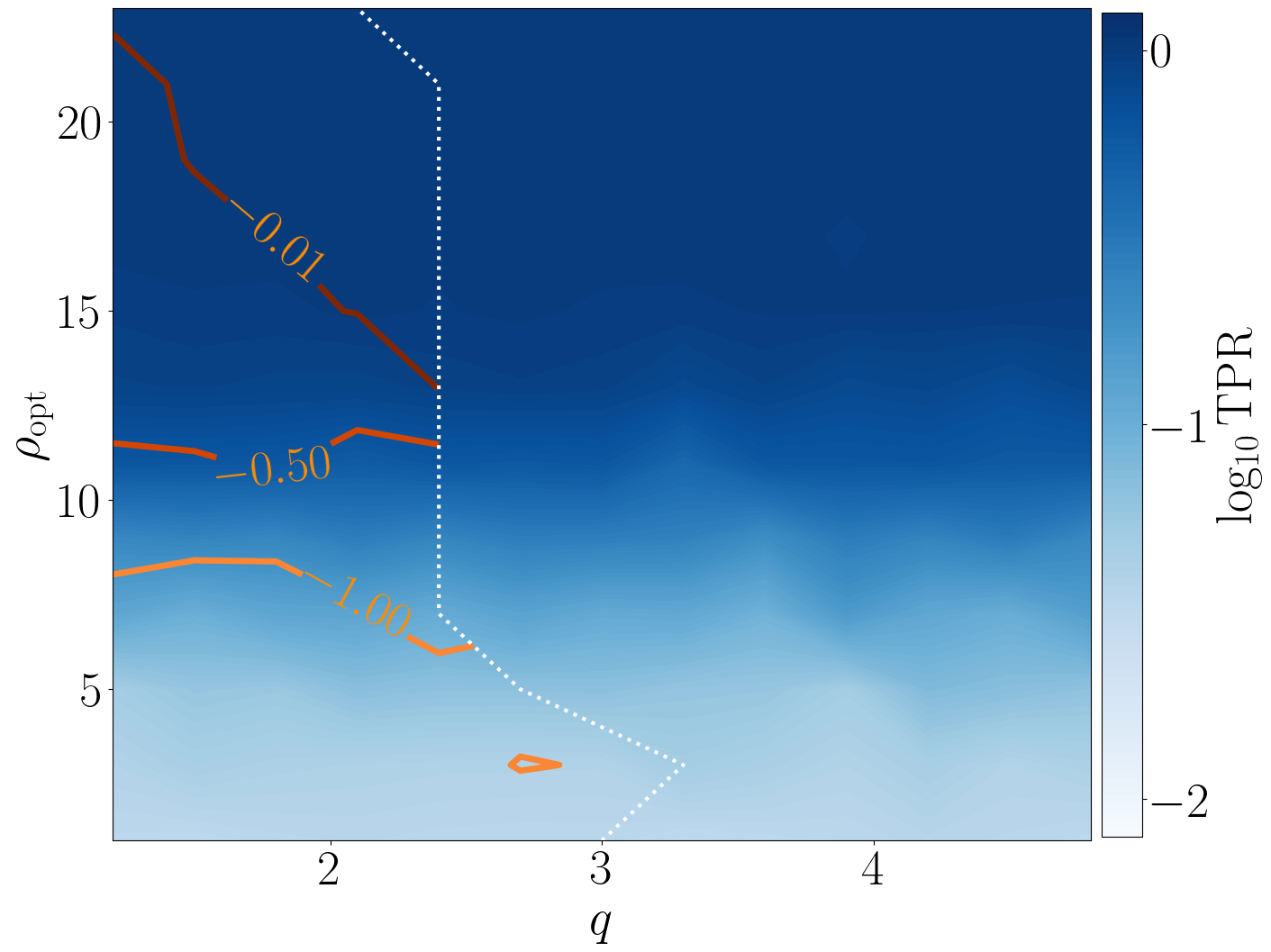}

    \caption{True positive rate of detection using the overlap metric and at a FAR of 12 $\rm{yr}^{-1}$ as a function of injected SNR $\rho_{\rm{opt}}$ versus varying primary black hole redshifted mass $\mathrm{M}_{1\,z}$ (top) and black hole mass ratio $q$ (bottom) in A+ sensitive LIGO-Hanford detector. The background colorplot represents the flat-prior test results and the iso-lines correspond to that of the astrophysical prior. The dotted white line represents the astrophysical population's upper limit, hence the astrophysical iso-lines terminate at this limit. 
    }
    \label{fig:TPR_minFAR_M1_q}
\end{figure}

To discriminate between the effects of the SNR and the BH redshifted mass on the detection efficiency, and to also look at its effect on the detection rates of the astrophysical population, we plot in Fig.~\ref{fig:TPR_minFAR_M1_q} the true positive rate of a BBH system as function of its optimal SNR $\rho_{\rm{opt}}$ and its redshifted primary mass $M_{1\,z}$ and mass ratio $q$. We observe that 
both the astrophysical and the flat-prior population have contours that remain relatively constant for $M_{1\,z} \geq 10 ~\rm{M_{\odot}}$. For smaller masses ($M_{1\,z} < 10 ~\rm{M_{\odot}}$), the efficiency of the flat-prior population drops due to the slightly diminishing performance of the SDC at higher GW frequencies. However, for the astrophysical population, this effect is counteracted by the large fraction of smaller black holes, resulting in a flat contour.

There is no preference for black hole mass ratio $q$; a true positive rate larger than $0.9$ is found for $\rho_{\rm{opt}} \geq 12.5$ for all mass ratios in the flat prior population. The best reconstructions in the astrophysical population is made for $q \sim 1$. These differences can be explained by the difference in studied populations and the trained population.

\subsection{Detection of O3b BBH signals}
\label{sec: O3bResults}

\begin{figure}
    \includegraphics[width=\linewidth]{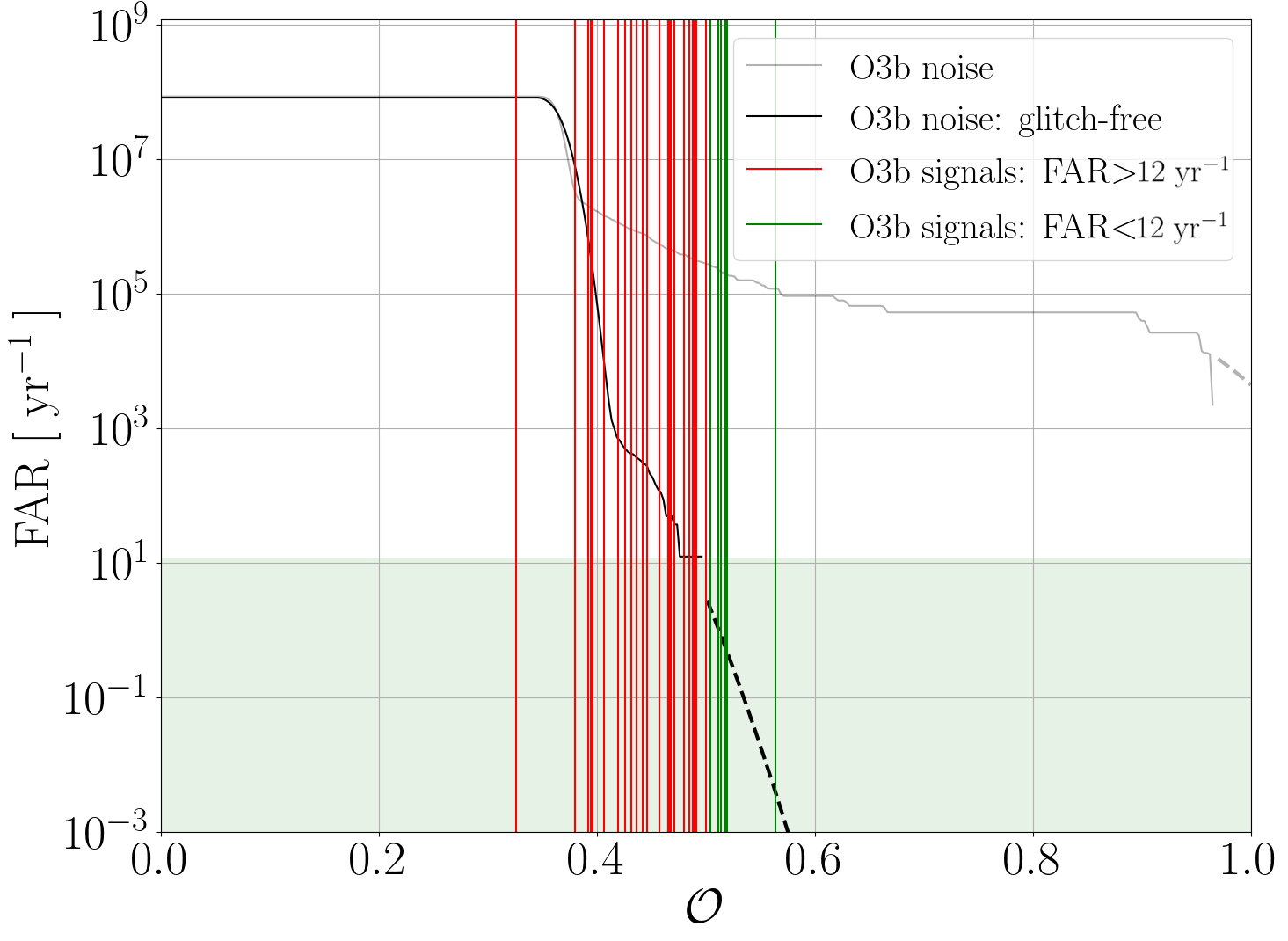}
    \caption{O3b confident event reconstruction performance in comparison to O3b noise before and after glitch removal using the overlap metric. The dashed lines represent the extrapolated FAR, and the shaded green region indicates FAR below 12 $\rm{yr}^{-1}$.}
    \label{fig:O3bResults}
\end{figure}

We next apply our algorithm to the 35 confident event detections from the O3b observing 
run~\cite{O3Search_GWOSC}. These are CBC events that have a probability of astrophysical origin $p_\mathrm{astro}>0.5$,  based upon results of at least one of the LVK search pipelines. For each confident event, the detector data is processed in the same manner as for the astrophysical population testing set. We reconstruct waveform $h_r$ from data $s$, and calculate the overlap.  
Approximately 2\% of the calculated noise overlap distribution is greater than 0.4 - some of which are larger than O3b confident reconstruction results. These are sourced from the reconstruction of transient noise, or glitches, in measured detector data. Noise transients clearly present an issue in real data analysis. We use the Gravity Spy glitch catalogue~\cite{glanzer_2021_5649212} to remove identified glitches in the selected O3b data and use this filtered noise data for our FAR analysis. 
In Fig.~\ref{fig:O3bResults} we plot the FAR curves of the noise distribution before and after glitch removal. As expected, the removal of glitch transients has a strong impact on detection prospects. 
O3b confident events that reconstructed with $\mathcal{O}(s, h_r) \geq 0.5$ had a FAR $\leq 12$ $\rm{yr}^{-1}$ - namely events GW191204\textunderscore171526, GW191230\textunderscore180458, GW200202\textunderscore154313, GW200216\textunderscore220804, GW200225\textunderscore060421 and GW200311\textunderscore115853. 
Reconstruction of all 35 confident events took 24.6~s total on an AMD EPYC 7502 CPU using the \texttt{time} Python software package. 
These results suggest that using SDL to reconstruct BBHs from LIGO data is a promising, expedient detection method when glitches have been removed from measured data. 

\section{Discussion and conclusions}
\label{sec:discussion} 

We have developed and tested a sparse dictionary learning (SDL) algorithm for the rapid detection of GWs from binary black hole mergers. Contrary to most detection pipelines of BBH systems currently operational in the LVK Collaboration, this approach does not rely on matched-filtering. 
To test our method, a suite of BBH systems from a realistic astrophysical population has been generated. Moreover, we have also built an additional set of BBH systems using a flat prior in order to test for potential biases in the design of the method. The SDL algorithm has been assessed using both, simulated data from the two BBH populations injected into the proposed A+ detector sensitivity {\it and} real data containing confident detections from the third LVK observing run. 

A few important comments on the design of a SDL pipeline for waveform reconstruction and detection are in order. When designing a dictionary, one must decide what features of a waveform to attempt to reconstruct. Although we chose here to reconstruct the inspiral, merger and ringdown of {\it all} BBH waveforms of our datasets over 0.375~s of observation time to create a {\it single}, general-purpose dictionary, one could instead create dictionaries to search for specific features over different time-windows. For example, one could create different dictionaries, each focusing on separately reconstructing the inspiral, merger and ringdown phases of the waveform in detail respectively, as opposed to a single dictionary that reconstructs the entire GW. 
Building dictionaries is strongly driven by the content of the datasets - ensuring one has a sufficient amount of quality training data is critical to signal reconstruction success. Dictionaries with too few training signals, or training signals that differ too greatly from the desired features to be reconstructed, may give poor results. Data streams with too small of a sampling frequency risk losing sinusoidal behaviors that can be reconstructed, we find that signals with a larger number of oscillation cycles are intrinsically more difficult to reconstruct with SDL. However, much of a BBH's signal power is found at frequencies below 1000 Hz, so oversampling will likely result in computational inefficiencies. More work is needed to understand a sufficient sampling rate for BBH waveform analysis in SDL contexts. Additionally, this study focuses on non-spinning BBHs - systems with spin adds additional effects on the resulting waveform's phasing~\cite{PhysRevD.74.041501}, an important consideration in the usage of SDL. Careful treatment by creating spin-aligned template banks are assembled for searches for this reason~\cite{Ajith:2009bn}. 

Considering too short of a time window may result in a partial or non-capture of a CBC waveform of interest; correspondingly, too long a time window may result in detector noise being weighted too much in detection statistics. Even with all of these considerations in mind, one may be designing dictionaries that cater to a subset of desired features while complicating the reconstruction capabilities of other features: formulating a general-purpose dictionary can thus become an intricate task when one wants to search for several waveform features. Once a learned dictionary has been created, verifying it is capable of reconstructing desired behaviors is necessary before using it on testing data. Searching for an optimal dictionary can be challenging - one needs to search through many combinations of hyperparameters numerically as no analytical approaches currently exist. This suggests that creating a truly optimised dictionary is a difficult task, but one can feasibly determine a justifiably good dictionary amongst a large set of them.

With these considerations in mind, we searched for a learned dictionary that reconstructs well the inspiral, merger and ringdown of a BBH wavefrom. We created 513 dictionaries varying in patch length $w$, number of patches $p$, regularisation parameter $\lambda$, and number of training waveforms and determined the quality of the learned dictionary using the overlap metric of reconstructed validation signals. We found that $w=512$, $p=768$, $\lambda=10^{-4}$ for a dictionary with 150 waveforms gave among the largest overlaps, and we fixed this dictionary for the study reported in this paper. Let us clarify that finding the best dictionary is beyond the scope of our study.

We find that the overlap metric is a better detection metric in comparison to the matched SNR metric. This is due to SDL's underestimation of amplitude in signal reconstruction, undercutting the matched SNR in comparison to the overlap which only takes phase similarities into account. Improvements on the SDL algorithm to better estimate amplitude could make the matched SNR statistic a more viable detection metric.

Using the LIGO Hanford detector at the A+ sensitivity, BBH signals from the astrophysical population with injected optimal SNR $\rho_{\rm{opt}} \geq 15$ are reconstructed with $\mathcal{O}(s, h_r) \geq 0.35$, corresponding to a $\rm{FAR} \leq \rm{12 \, yr}^{-1}$. The best reconstruction performance is found for BBH systems with primary mass $M_1 \sim 40\,\rm{M}_\odot$ and $q \sim 1$ - such systems with $\rho_{\rm{opt}} \geq 12.5$ could be detected with a true positive rate  $\geq 0.9$ when $\rm{FAR = 12\, yr}^{-1}$. The flat prior results proved to have equal detection prospects for systems with $M_{\rm{tot}} \geq 20\,\rm{M}_\odot$ and $q \geq 1$, only breaking down at smaller masses and mass ratio due to limitations in the waveform frequency content of the dictionary in such cases. This suggests that detection prospects using SDL, although promising, are determined by dictionary and testing population designs. We have also shown that applying SDL to actual data instead of simulated signals, using in particular the 35 BBH confident detections from O3b, yielded reasonably good reconstruction performance as measured by the overlap. Glitches have a strong impact on the detection prospects when working with real data, although making up a small portion of detector noise data. 
A removal of glitches using the noise transient catalogue from~\cite{glanzer_2021_5649212} showed a dramatic improvement in detection prospects, with any O3b confident result reconstructed with $\mathcal{O} \geq 0.5$ having a FAR = $12\,\rm{yr}^{-1}$.

Perhaps most importantly, SDL has also proven itself to be a fairly rapidly working method. With an assembled dictionary, one can reconstruct a single testing signal within seconds - allowing for potentially rapid analysis of a GW data stream. Learning a dictionary with the 150 training signals from our BBH populations (with uniformly distributed primary masses, sky location, and redshift) took, at most, 454~s to create, and reconstruction of all 35 O3b confident events took 24.6~s - potentially faster than traditional matched-filtering pipelines, depending on the set-up of the analysis. However, an explicit computational speed study needs to be conducted as this comparison is not trivial.

We are beginning to scratch the surface of machine learning applications in GW science. They have shown to be able to yield improved GW detection, but also have the potential to search for atypical or difficult to model GW signals such as bursts~\cite{Adams_2013, abbott2019search, Kibble_1976, abbott2018constraints}. An improvement in analysis quality can also allow for greater clarity on how more subtle effects of precession and eccentricity can affect detection prospects. As we collect greater volumes of observational data with additional observation runs including additional detector baselines, accelerated machine learning algorithms can allow efficient processing and analysis of such data.

In comparison to traditional search methods, a matched filter template-based search is expected to outperform SDL in its present form. We find that the universe volume ratio between the two methods are $0.81$ and $0.66$ for the simulated astrophysical and flat-prior datasets. Future study aims to expand the single-detector SDL method reported in this work to multiple detectors for a network reconstruction analysis. The inclusion of multiple streams of data to reconstruct can improve detection prospects and allow for better comparisons to other detection pipelines. The additional detectors can also provide an improved error analysis: coincidence rates between detectors can be determined, usually reducing the FAR by orders of magnitude. SDL's expedient analysis speed could potentially be used in as a pre-screening tool in LVK data. Further work is needed  to separate overlapping CBC signals, and study the viability of subtracting reconstructed BBHs from ground detector data in an effort to detect or constrain weaker GW sources - namely from cosmological origin. More developed search methods for an optimised dictionary are needed as well in order to improve dictionary-learning detection approaches. Furthermore, the inclusion of additional noise sources may introduce degeneracies when reconstructing CBCs - we leave the impact of this on our analysis for future study.

{\bf Data availability:} The codes and all the necessary files to reproduce the results in this paper will be made available on GitHub\footnote{https://github.com/Rahul-Srinivasan/GW\_DictionaryLearning}.

\acknowledgments{
We thank Juan Calderon Bustillo for the useful discussions on search pipeline comparisons. This material is based upon work supported by NSF's LIGO Laboratory which is a major facility fully funded by the National Science Foundation. We acknowledge computational resources provided by the LIGO Laboratory and supported by National Science Foundation Grants No. PHY-0757058 and No. PHY-0823459. The software packages used in this study are \texttt{matplotlib} ~\cite{matplotlib}, \texttt{numpy} ~\cite{numpy}, \texttt{PyCBC} ~\cite{alex_nitz_2024_10473621} and \texttt{HTCondor} ~\cite{condor}.
JAF and ATF are supported by the Spanish Agencia Estatal de Investigaci\'on (grant PID2021-125485NB-C21) funded by MCIN/AEI/10.13039/501100011033 and ERDF A way of making Europe, by the Generalitat Valenciana (grant CIPROM/2022/49), and by the European Horizon Europe staff exchange (SE) programme HORIZON-MSCA-2021-SE-01 (NewFunFiCo-101086251). The work of MS is partially supported by the Science and Technology Facilities Council (STFC grant ST/X000753/1). AL is supported by the ANR COSMERGE project, grant ANR-20-CE31-001 of the French Agence Nationale de la Recherche. RS and AL acknowledge support from the graduate and research school EUR SPECTRUM. RS is also supported by the European Union’s H2020 ERC Consolidator Grant ``GRavity from Astrophysical to Microscopic Scales'' (Grant No. GRAMS-815673), the PRIN 2022 grant ``GUVIRP - Gravity tests in the UltraViolet and InfraRed with Pulsar timing'', and the EU Horizon 2020 Research and Innovation Programme under the Marie Sklodowska-Curie Grant Agreement No.~101007855. 

This manuscript was assigned LIGO-Document number  LIGO-P2400194.
}

\bibliography{main_restructured}

\begin{thebibliography}{58}%
\makeatletter
\providecommand \@ifxundefined [1]{%
 \@ifx{#1\undefined}
}%
\providecommand \@ifnum [1]{%
 \ifnum #1\expandafter \@firstoftwo
 \else \expandafter \@secondoftwo
 \fi
}%
\providecommand \@ifx [1]{%
 \ifx #1\expandafter \@firstoftwo
 \else \expandafter \@secondoftwo
 \fi
}%
\providecommand \natexlab [1]{#1}%
\providecommand \enquote  [1]{``#1''}%
\providecommand \bibnamefont  [1]{#1}%
\providecommand \bibfnamefont [1]{#1}%
\providecommand \citenamefont [1]{#1}%
\providecommand \href@noop [0]{\@secondoftwo}%
\providecommand \href [0]{\begingroup \@sanitize@url \@href}%
\providecommand \@href[1]{\@@startlink{#1}\@@href}%
\providecommand \@@href[1]{\endgroup#1\@@endlink}%
\providecommand \@sanitize@url [0]{\catcode `\\12\catcode `\$12\catcode `\&12\catcode `\#12\catcode `\^12\catcode `\_12\catcode `\%12\relax}%
\providecommand \@@startlink[1]{}%
\providecommand \@@endlink[0]{}%
\providecommand \url  [0]{\begingroup\@sanitize@url \@url }%
\providecommand \@url [1]{\endgroup\@href {#1}{\urlprefix }}%
\providecommand \urlprefix  [0]{URL }%
\providecommand \Eprint [0]{\href }%
\providecommand \doibase [0]{http://dx.doi.org/}%
\providecommand \selectlanguage [0]{\@gobble}%
\providecommand \bibinfo  [0]{\@secondoftwo}%
\providecommand \bibfield  [0]{\@secondoftwo}%
\providecommand \translation [1]{[#1]}%
\providecommand \BibitemOpen [0]{}%
\providecommand \bibitemStop [0]{}%
\providecommand \bibitemNoStop [0]{.\EOS\space}%
\providecommand \EOS [0]{\spacefactor3000\relax}%
\providecommand \BibitemShut  [1]{\csname bibitem#1\endcsname}%
\let\auto@bib@innerbib\@empty
\bibitem [{\citenamefont {Abbott}\ \emph {et~al.}(2016)\citenamefont {Abbott} \emph {et~al.}}]{FirstGWDetec}%
  \BibitemOpen
  \bibfield  {author} {\bibinfo {author} {\bibfnamefont {B.~P.}\ \bibnamefont {Abbott}} \emph {et~al.} (\bibinfo {collaboration} {LIGO Scientific, Virgo}),\ }\href {\doibase 10.1103/PhysRevLett.116.061102} {\bibfield  {journal} {\bibinfo  {journal} {Phys. Rev. Lett.}\ }\textbf {\bibinfo {volume} {116}},\ \bibinfo {pages} {061102} (\bibinfo {year} {2016})},\ \Eprint {http://arxiv.org/abs/1602.03837} {arXiv:1602.03837 [gr-qc]} \BibitemShut {NoStop}%
\bibitem [{\citenamefont {Aasi}\ \emph {et~al.}(2015)\citenamefont {Aasi} \emph {et~al.}}]{Aasi:2014jea}%
  \BibitemOpen
  \bibfield  {author} {\bibinfo {author} {\bibfnamefont {J.}~\bibnamefont {Aasi}} \emph {et~al.} (\bibinfo {collaboration} {LIGO Scientific Collaboration}),\ }\href {\doibase 10.1088/0264-9381/32/7/074001} {\bibfield  {journal} {\bibinfo  {journal} {Class. Quant. Grav.}\ }\textbf {\bibinfo {volume} {32}},\ \bibinfo {pages} {074001} (\bibinfo {year} {2015})},\ \Eprint {http://arxiv.org/abs/1411.4547} {arXiv:1411.4547 [gr-qc]} \BibitemShut {NoStop}%
\bibitem [{\citenamefont {Acernese}\ \emph {et~al.}(2015)\citenamefont {Acernese} \emph {et~al.}}]{Acernese:2014hva}%
  \BibitemOpen
  \bibfield  {author} {\bibinfo {author} {\bibfnamefont {F.}~\bibnamefont {Acernese}} \emph {et~al.} (\bibinfo {collaboration} {Virgo Collaboration}),\ }\href {\doibase 10.1088/0264-9381/32/2/024001} {\bibfield  {journal} {\bibinfo  {journal} {Class. Quant. Grav.}\ }\textbf {\bibinfo {volume} {32}},\ \bibinfo {pages} {024001} (\bibinfo {year} {2015})},\ \Eprint {http://arxiv.org/abs/1408.3978} {arXiv:1408.3978 [gr-qc]} \BibitemShut {NoStop}%
\bibitem [{\citenamefont {Abbott}\ \emph {et~al.}(2021)\citenamefont {Abbott} \emph {et~al.}}]{O1O2Search_GWOSC}%
  \BibitemOpen
  \bibfield  {author} {\bibinfo {author} {\bibfnamefont {R.}~\bibnamefont {Abbott}} \emph {et~al.} (\bibinfo {collaboration} {LIGO Scientific, Virgo}),\ }\href {\doibase 10.1016/j.softx.2021.100658} {\bibfield  {journal} {\bibinfo  {journal} {SoftwareX}\ }\textbf {\bibinfo {volume} {13}},\ \bibinfo {pages} {100658} (\bibinfo {year} {2021})},\ \Eprint {http://arxiv.org/abs/1912.11716} {arXiv:1912.11716 [gr-qc]} \BibitemShut {NoStop}%
\bibitem [{\citenamefont {Abbott}\ \emph {et~al.}(2023)\citenamefont {Abbott} \emph {et~al.}}]{O3Search_GWOSC}%
  \BibitemOpen
  \bibfield  {author} {\bibinfo {author} {\bibfnamefont {R.}~\bibnamefont {Abbott}} \emph {et~al.},\ }\href {\doibase 10.3847/1538-4365/acdc9f} {\bibfield  {journal} {\bibinfo  {journal} {The Astrophysical Journal Supplement Series}\ }\textbf {\bibinfo {volume} {267}},\ \bibinfo {pages} {29} (\bibinfo {year} {2023})}\BibitemShut {NoStop}%
\bibitem [{\citenamefont {Akutsu}\ \emph {et~al.}(2021)\citenamefont {Akutsu} \emph {et~al.}}]{KAGRA:2020tym}%
  \BibitemOpen
  \bibfield  {author} {\bibinfo {author} {\bibfnamefont {T.}~\bibnamefont {Akutsu}} \emph {et~al.} (\bibinfo {collaboration} {KAGRA}),\ }\href {\doibase 10.1093/ptep/ptaa125} {\bibfield  {journal} {\bibinfo  {journal} {PTEP}\ }\textbf {\bibinfo {volume} {2021}},\ \bibinfo {pages} {05A101} (\bibinfo {year} {2021})},\ \Eprint {http://arxiv.org/abs/2005.05574} {arXiv:2005.05574 [physics.ins-det]} \BibitemShut {NoStop}%
\bibitem [{\citenamefont {{Abbott}}\ \emph {et~al.}(2023)\citenamefont {{Abbott}} \emph {et~al.}}]{GWTC-3_Population_Properties}%
  \BibitemOpen
  \bibfield  {author} {\bibinfo {author} {\bibfnamefont {R.}~\bibnamefont {{Abbott}}} \emph {et~al.},\ }\href {\doibase 10.1103/PhysRevX.13.011048} {\bibfield  {journal} {\bibinfo  {journal} {Physical Review X}\ }\textbf {\bibinfo {volume} {13}},\ \bibinfo {eid} {011048} (\bibinfo {year} {2023})},\ \Eprint {http://arxiv.org/abs/2111.03634} {arXiv:2111.03634 [astro-ph.HE]} \BibitemShut {NoStop}%
\bibitem [{\citenamefont {Sachdev}\ \emph {et~al.}(2019)\citenamefont {Sachdev}, \citenamefont {Caudill}, \citenamefont {Fong}, \citenamefont {Lo}, \citenamefont {Messick}, \citenamefont {Mukherjee}, \citenamefont {Magee}, \citenamefont {Tsukada}, \citenamefont {Blackburn}, \citenamefont {Brady}, \citenamefont {Brockill}, \citenamefont {Cannon}, \citenamefont {Chamberlin}, \citenamefont {Chatterjee}, \citenamefont {Creighton}, \citenamefont {Godwin}, \citenamefont {Gupta}, \citenamefont {Hanna}, \citenamefont {Kapadia}, \citenamefont {Lang}, \citenamefont {Li}, \citenamefont {Meacher}, \citenamefont {Pace}, \citenamefont {Privitera}, \citenamefont {Sadeghian}, \citenamefont {Wade}, \citenamefont {Wade}, \citenamefont {Weinstein},\ and\ \citenamefont {Xiao}}]{sachdev2019gstlal}%
  \BibitemOpen
  \bibfield  {author} {\bibinfo {author} {\bibfnamefont {S.}~\bibnamefont {Sachdev}}, \bibinfo {author} {\bibfnamefont {S.}~\bibnamefont {Caudill}}, \bibinfo {author} {\bibfnamefont {H.}~\bibnamefont {Fong}}, \bibinfo {author} {\bibfnamefont {R.~K.~L.}\ \bibnamefont {Lo}}, \bibinfo {author} {\bibfnamefont {C.}~\bibnamefont {Messick}}, \bibinfo {author} {\bibfnamefont {D.}~\bibnamefont {Mukherjee}}, \bibinfo {author} {\bibfnamefont {R.}~\bibnamefont {Magee}}, \bibinfo {author} {\bibfnamefont {L.}~\bibnamefont {Tsukada}}, \bibinfo {author} {\bibfnamefont {K.}~\bibnamefont {Blackburn}}, \bibinfo {author} {\bibfnamefont {P.}~\bibnamefont {Brady}}, \bibinfo {author} {\bibfnamefont {P.}~\bibnamefont {Brockill}}, \bibinfo {author} {\bibfnamefont {K.}~\bibnamefont {Cannon}}, \bibinfo {author} {\bibfnamefont {S.~J.}\ \bibnamefont {Chamberlin}}, \bibinfo {author} {\bibfnamefont {D.}~\bibnamefont {Chatterjee}}, \bibinfo {author} {\bibfnamefont {J.~D.~E.}\ \bibnamefont {Creighton}}, \bibinfo {author} {\bibfnamefont
  {P.}~\bibnamefont {Godwin}}, \bibinfo {author} {\bibfnamefont {A.}~\bibnamefont {Gupta}}, \bibinfo {author} {\bibfnamefont {C.}~\bibnamefont {Hanna}}, \bibinfo {author} {\bibfnamefont {S.}~\bibnamefont {Kapadia}}, \bibinfo {author} {\bibfnamefont {R.~N.}\ \bibnamefont {Lang}}, \bibinfo {author} {\bibfnamefont {T.~G.~F.}\ \bibnamefont {Li}}, \bibinfo {author} {\bibfnamefont {D.}~\bibnamefont {Meacher}}, \bibinfo {author} {\bibfnamefont {A.}~\bibnamefont {Pace}}, \bibinfo {author} {\bibfnamefont {S.}~\bibnamefont {Privitera}}, \bibinfo {author} {\bibfnamefont {L.}~\bibnamefont {Sadeghian}}, \bibinfo {author} {\bibfnamefont {L.}~\bibnamefont {Wade}}, \bibinfo {author} {\bibfnamefont {M.}~\bibnamefont {Wade}}, \bibinfo {author} {\bibfnamefont {A.}~\bibnamefont {Weinstein}}, \ and\ \bibinfo {author} {\bibfnamefont {S.~L.}\ \bibnamefont {Xiao}},\ }\href@noop {} {\enquote {\bibinfo {title} {The gstlal search analysis methods for compact binary mergers in advanced ligo's second and advanced virgo's first observing
  runs},}\ } (\bibinfo {year} {2019}),\ \Eprint {http://arxiv.org/abs/1901.08580} {arXiv:1901.08580 [gr-qc]} \BibitemShut {NoStop}%
\bibitem [{\citenamefont {Aubin}\ \emph {et~al.}(2021)\citenamefont {Aubin}, \citenamefont {Brighenti}, \citenamefont {Chierici}, \citenamefont {Estevez}, \citenamefont {Greco}, \citenamefont {Guidi}, \citenamefont {Juste}, \citenamefont {Marion}, \citenamefont {Mours}, \citenamefont {Nitoglia}, \citenamefont {Sauter},\ and\ \citenamefont {Sordini}}]{Aubin_2021}%
  \BibitemOpen
  \bibfield  {author} {\bibinfo {author} {\bibfnamefont {F.}~\bibnamefont {Aubin}}, \bibinfo {author} {\bibfnamefont {F.}~\bibnamefont {Brighenti}}, \bibinfo {author} {\bibfnamefont {R.}~\bibnamefont {Chierici}}, \bibinfo {author} {\bibfnamefont {D.}~\bibnamefont {Estevez}}, \bibinfo {author} {\bibfnamefont {G.}~\bibnamefont {Greco}}, \bibinfo {author} {\bibfnamefont {G.~M.}\ \bibnamefont {Guidi}}, \bibinfo {author} {\bibfnamefont {V.}~\bibnamefont {Juste}}, \bibinfo {author} {\bibfnamefont {F.}~\bibnamefont {Marion}}, \bibinfo {author} {\bibfnamefont {B.}~\bibnamefont {Mours}}, \bibinfo {author} {\bibfnamefont {E.}~\bibnamefont {Nitoglia}}, \bibinfo {author} {\bibfnamefont {O.}~\bibnamefont {Sauter}}, \ and\ \bibinfo {author} {\bibfnamefont {V.}~\bibnamefont {Sordini}},\ }\href {\doibase 10.1088/1361-6382/abe913} {\bibfield  {journal} {\bibinfo  {journal} {Classical and Quantum Gravity}\ }\textbf {\bibinfo {volume} {38}},\ \bibinfo {pages} {095004} (\bibinfo {year} {2021})}\BibitemShut {NoStop}%
\bibitem [{\citenamefont {Canton}\ \emph {et~al.}(2021)\citenamefont {Canton}, \citenamefont {Nitz}, \citenamefont {Gadre}, \citenamefont {Davies}, \citenamefont {Villa-Ortega}, \citenamefont {Dent}, \citenamefont {Harry},\ and\ \citenamefont {Xiao}}]{Dal_Canton_2021}%
  \BibitemOpen
  \bibfield  {author} {\bibinfo {author} {\bibfnamefont {T.~D.}\ \bibnamefont {Canton}}, \bibinfo {author} {\bibfnamefont {A.~H.}\ \bibnamefont {Nitz}}, \bibinfo {author} {\bibfnamefont {B.}~\bibnamefont {Gadre}}, \bibinfo {author} {\bibfnamefont {G.~S.~C.}\ \bibnamefont {Davies}}, \bibinfo {author} {\bibfnamefont {V.}~\bibnamefont {Villa-Ortega}}, \bibinfo {author} {\bibfnamefont {T.}~\bibnamefont {Dent}}, \bibinfo {author} {\bibfnamefont {I.}~\bibnamefont {Harry}}, \ and\ \bibinfo {author} {\bibfnamefont {L.}~\bibnamefont {Xiao}},\ }\href {\doibase 10.3847/1538-4357/ac2f9a} {\bibfield  {journal} {\bibinfo  {journal} {The Astrophysical Journal}\ }\textbf {\bibinfo {volume} {923}},\ \bibinfo {pages} {254} (\bibinfo {year} {2021})}\BibitemShut {NoStop}%
\bibitem [{\citenamefont {Venumadhav}\ \emph {et~al.}(2020)\citenamefont {Venumadhav}, \citenamefont {Zackay}, \citenamefont {Roulet}, \citenamefont {Dai},\ and\ \citenamefont {Zaldarriaga}}]{Venumadhav:2019lyq}%
  \BibitemOpen
  \bibfield  {author} {\bibinfo {author} {\bibfnamefont {T.}~\bibnamefont {Venumadhav}}, \bibinfo {author} {\bibfnamefont {B.}~\bibnamefont {Zackay}}, \bibinfo {author} {\bibfnamefont {J.}~\bibnamefont {Roulet}}, \bibinfo {author} {\bibfnamefont {L.}~\bibnamefont {Dai}}, \ and\ \bibinfo {author} {\bibfnamefont {M.}~\bibnamefont {Zaldarriaga}},\ }\href {\doibase 10.1103/PhysRevD.101.083030} {\bibfield  {journal} {\bibinfo  {journal} {Phys. Rev. D}\ }\textbf {\bibinfo {volume} {101}},\ \bibinfo {pages} {083030} (\bibinfo {year} {2020})},\ \Eprint {http://arxiv.org/abs/1904.07214} {arXiv:1904.07214 [astro-ph.HE]} \BibitemShut {NoStop}%
\bibitem [{\citenamefont {Chu}\ \emph {et~al.}(2021)\citenamefont {Chu}, \citenamefont {Kovalam}, \citenamefont {Wen}, \citenamefont {Slaven-Blair}, \citenamefont {Bosveld}, \citenamefont {Chen}, \citenamefont {Clearwater}, \citenamefont {Codoreanu}, \citenamefont {Du}, \citenamefont {Guo}, \citenamefont {Guo}, \citenamefont {Kim}, \citenamefont {Li}, \citenamefont {Oloworaran}, \citenamefont {Panther}, \citenamefont {Powell}, \citenamefont {Sengupta}, \citenamefont {Wette},\ and\ \citenamefont {Zhu}}]{chu2021spiir}%
  \BibitemOpen
  \bibfield  {author} {\bibinfo {author} {\bibfnamefont {Q.}~\bibnamefont {Chu}}, \bibinfo {author} {\bibfnamefont {M.}~\bibnamefont {Kovalam}}, \bibinfo {author} {\bibfnamefont {L.}~\bibnamefont {Wen}}, \bibinfo {author} {\bibfnamefont {T.}~\bibnamefont {Slaven-Blair}}, \bibinfo {author} {\bibfnamefont {J.}~\bibnamefont {Bosveld}}, \bibinfo {author} {\bibfnamefont {Y.}~\bibnamefont {Chen}}, \bibinfo {author} {\bibfnamefont {P.}~\bibnamefont {Clearwater}}, \bibinfo {author} {\bibfnamefont {A.}~\bibnamefont {Codoreanu}}, \bibinfo {author} {\bibfnamefont {Z.}~\bibnamefont {Du}}, \bibinfo {author} {\bibfnamefont {X.}~\bibnamefont {Guo}}, \bibinfo {author} {\bibfnamefont {X.}~\bibnamefont {Guo}}, \bibinfo {author} {\bibfnamefont {K.}~\bibnamefont {Kim}}, \bibinfo {author} {\bibfnamefont {T.~G.~F.}\ \bibnamefont {Li}}, \bibinfo {author} {\bibfnamefont {V.}~\bibnamefont {Oloworaran}}, \bibinfo {author} {\bibfnamefont {F.}~\bibnamefont {Panther}}, \bibinfo {author} {\bibfnamefont {J.}~\bibnamefont {Powell}},
  \bibinfo {author} {\bibfnamefont {A.~S.}\ \bibnamefont {Sengupta}}, \bibinfo {author} {\bibfnamefont {K.}~\bibnamefont {Wette}}, \ and\ \bibinfo {author} {\bibfnamefont {X.}~\bibnamefont {Zhu}},\ }\href@noop {} {\enquote {\bibinfo {title} {The spiir online coherent pipeline to search for gravitational waves from compact binary coalescences},}\ } (\bibinfo {year} {2021}),\ \Eprint {http://arxiv.org/abs/2011.06787} {arXiv:2011.06787 [gr-qc]} \BibitemShut {NoStop}%
\bibitem [{\citenamefont {Klimenko}\ \emph {et~al.}(2016)\citenamefont {Klimenko}, \citenamefont {Vedovato}, \citenamefont {Drago}, \citenamefont {Salemi}, \citenamefont {Tiwari}, \citenamefont {Prodi}, \citenamefont {Lazzaro}, \citenamefont {Ackley}, \citenamefont {Tiwari}, \citenamefont {Da~Silva},\ and\ \citenamefont {Mitselmakher}}]{Klimenko_2016}%
  \BibitemOpen
  \bibfield  {author} {\bibinfo {author} {\bibfnamefont {S.}~\bibnamefont {Klimenko}}, \bibinfo {author} {\bibfnamefont {G.}~\bibnamefont {Vedovato}}, \bibinfo {author} {\bibfnamefont {M.}~\bibnamefont {Drago}}, \bibinfo {author} {\bibfnamefont {F.}~\bibnamefont {Salemi}}, \bibinfo {author} {\bibfnamefont {V.}~\bibnamefont {Tiwari}}, \bibinfo {author} {\bibfnamefont {G.}~\bibnamefont {Prodi}}, \bibinfo {author} {\bibfnamefont {C.}~\bibnamefont {Lazzaro}}, \bibinfo {author} {\bibfnamefont {K.}~\bibnamefont {Ackley}}, \bibinfo {author} {\bibfnamefont {S.}~\bibnamefont {Tiwari}}, \bibinfo {author} {\bibfnamefont {C.}~\bibnamefont {Da~Silva}}, \ and\ \bibinfo {author} {\bibfnamefont {G.}~\bibnamefont {Mitselmakher}},\ }\href {\doibase 10.1103/physrevd.93.042004} {\bibfield  {journal} {\bibinfo  {journal} {Physical Review D}\ }\textbf {\bibinfo {volume} {93}} (\bibinfo {year} {2016}),\ 10.1103/physrevd.93.042004}\BibitemShut {NoStop}%
\bibitem [{\citenamefont {Abbott}\ \emph {et~al.}(2020)\citenamefont {Abbott} \emph {et~al.}}]{LIGOScientific:2019hgc}%
  \BibitemOpen
  \bibfield  {author} {\bibinfo {author} {\bibfnamefont {B.~P.}\ \bibnamefont {Abbott}} \emph {et~al.} (\bibinfo {collaboration} {LIGO Scientific, Virgo}),\ }\href {\doibase 10.1088/1361-6382/ab685e} {\bibfield  {journal} {\bibinfo  {journal} {Class. Quant. Grav.}\ }\textbf {\bibinfo {volume} {37}},\ \bibinfo {pages} {055002} (\bibinfo {year} {2020})},\ \Eprint {http://arxiv.org/abs/1908.11170} {arXiv:1908.11170 [gr-qc]} \BibitemShut {NoStop}%
\bibitem [{\citenamefont {Kim}\ \emph {et~al.}(2015)\citenamefont {Kim}, \citenamefont {Harry}, \citenamefont {Hodge}, \citenamefont {Kim}, \citenamefont {Lee}, \citenamefont {Lee}, \citenamefont {Oh}, \citenamefont {Oh},\ and\ \citenamefont {Son}}]{Kim_2015}%
  \BibitemOpen
  \bibfield  {author} {\bibinfo {author} {\bibfnamefont {K.}~\bibnamefont {Kim}}, \bibinfo {author} {\bibfnamefont {I.~W.}\ \bibnamefont {Harry}}, \bibinfo {author} {\bibfnamefont {K.~A.}\ \bibnamefont {Hodge}}, \bibinfo {author} {\bibfnamefont {Y.-M.}\ \bibnamefont {Kim}}, \bibinfo {author} {\bibfnamefont {C.-H.}\ \bibnamefont {Lee}}, \bibinfo {author} {\bibfnamefont {H.~K.}\ \bibnamefont {Lee}}, \bibinfo {author} {\bibfnamefont {J.~J.}\ \bibnamefont {Oh}}, \bibinfo {author} {\bibfnamefont {S.~H.}\ \bibnamefont {Oh}}, \ and\ \bibinfo {author} {\bibfnamefont {E.~J.}\ \bibnamefont {Son}},\ }\href {\doibase 10.1088/0264-9381/32/24/245002} {\bibfield  {journal} {\bibinfo  {journal} {Classical and Quantum Gravity}\ }\textbf {\bibinfo {volume} {32}},\ \bibinfo {pages} {245002} (\bibinfo {year} {2015})}\BibitemShut {NoStop}%
\bibitem [{\citenamefont {George}\ and\ \citenamefont {Huerta}(2018)}]{George_2018}%
  \BibitemOpen
  \bibfield  {author} {\bibinfo {author} {\bibfnamefont {D.}~\bibnamefont {George}}\ and\ \bibinfo {author} {\bibfnamefont {E.}~\bibnamefont {Huerta}},\ }\href {\doibase 10.1016/j.physletb.2017.12.053} {\bibfield  {journal} {\bibinfo  {journal} {Physics Letters B}\ }\textbf {\bibinfo {volume} {778}},\ \bibinfo {pages} {64–70} (\bibinfo {year} {2018})}\BibitemShut {NoStop}%
\bibitem [{\citenamefont {{Marx}}\ \emph {et~al.}(2024)\citenamefont {{Marx}}, \citenamefont {{Benoit}}, \citenamefont {{Gunny}}, \citenamefont {{Omer}}, \citenamefont {{Chatterjee}}, \citenamefont {{Venterea}}, \citenamefont {{Wills}}, \citenamefont {{Saleem}}, \citenamefont {{Moreno}}, \citenamefont {{Raikman}}, \citenamefont {{Govorkova}}, \citenamefont {{Rankin}}, \citenamefont {{Coughlin}}, \citenamefont {{Harris}},\ and\ \citenamefont {{Katsavounidis}}}]{Marx:2024}%
  \BibitemOpen
  \bibfield  {author} {\bibinfo {author} {\bibfnamefont {E.}~\bibnamefont {{Marx}}}, \bibinfo {author} {\bibfnamefont {W.}~\bibnamefont {{Benoit}}}, \bibinfo {author} {\bibfnamefont {A.}~\bibnamefont {{Gunny}}}, \bibinfo {author} {\bibfnamefont {R.}~\bibnamefont {{Omer}}}, \bibinfo {author} {\bibfnamefont {D.}~\bibnamefont {{Chatterjee}}}, \bibinfo {author} {\bibfnamefont {R.~C.}\ \bibnamefont {{Venterea}}}, \bibinfo {author} {\bibfnamefont {L.}~\bibnamefont {{Wills}}}, \bibinfo {author} {\bibfnamefont {M.}~\bibnamefont {{Saleem}}}, \bibinfo {author} {\bibfnamefont {E.}~\bibnamefont {{Moreno}}}, \bibinfo {author} {\bibfnamefont {R.}~\bibnamefont {{Raikman}}}, \bibinfo {author} {\bibfnamefont {E.}~\bibnamefont {{Govorkova}}}, \bibinfo {author} {\bibfnamefont {D.}~\bibnamefont {{Rankin}}}, \bibinfo {author} {\bibfnamefont {M.~W.}\ \bibnamefont {{Coughlin}}}, \bibinfo {author} {\bibfnamefont {P.}~\bibnamefont {{Harris}}}, \ and\ \bibinfo {author} {\bibfnamefont {E.}~\bibnamefont {{Katsavounidis}}},\ }\href
  {\doibase 10.48550/arXiv.2403.18661} {\bibfield  {journal} {\bibinfo  {journal} {arXiv e-prints}\ ,\ \bibinfo {eid} {arXiv:2403.18661}} (\bibinfo {year} {2024})},\ \Eprint {http://arxiv.org/abs/2403.18661} {arXiv:2403.18661 [gr-qc]} \BibitemShut {NoStop}%
\bibitem [{\citenamefont {Cuoco}\ \emph {et~al.}(2020)\citenamefont {Cuoco}, \citenamefont {Powell}, \citenamefont {Cavaglià}, \citenamefont {Ackley}, \citenamefont {Bejger}, \citenamefont {Chatterjee}, \citenamefont {Coughlin}, \citenamefont {Coughlin}, \citenamefont {Easter}, \citenamefont {Essick}, \citenamefont {Gabbard}, \citenamefont {Gebhard}, \citenamefont {Ghosh}, \citenamefont {Haegel}, \citenamefont {Iess}, \citenamefont {Keitel}, \citenamefont {Márka}, \citenamefont {Márka}, \citenamefont {Morawski}, \citenamefont {Nguyen}, \citenamefont {Ormiston}, \citenamefont {Pürrer}, \citenamefont {Razzano}, \citenamefont {Staats}, \citenamefont {Vajente},\ and\ \citenamefont {Williams}}]{Cuoco_2020}%
  \BibitemOpen
  \bibfield  {author} {\bibinfo {author} {\bibfnamefont {E.}~\bibnamefont {Cuoco}}, \bibinfo {author} {\bibfnamefont {J.}~\bibnamefont {Powell}}, \bibinfo {author} {\bibfnamefont {M.}~\bibnamefont {Cavaglià}}, \bibinfo {author} {\bibfnamefont {K.}~\bibnamefont {Ackley}}, \bibinfo {author} {\bibfnamefont {M.}~\bibnamefont {Bejger}}, \bibinfo {author} {\bibfnamefont {C.}~\bibnamefont {Chatterjee}}, \bibinfo {author} {\bibfnamefont {M.}~\bibnamefont {Coughlin}}, \bibinfo {author} {\bibfnamefont {S.}~\bibnamefont {Coughlin}}, \bibinfo {author} {\bibfnamefont {P.}~\bibnamefont {Easter}}, \bibinfo {author} {\bibfnamefont {R.}~\bibnamefont {Essick}}, \bibinfo {author} {\bibfnamefont {H.}~\bibnamefont {Gabbard}}, \bibinfo {author} {\bibfnamefont {T.}~\bibnamefont {Gebhard}}, \bibinfo {author} {\bibfnamefont {S.}~\bibnamefont {Ghosh}}, \bibinfo {author} {\bibfnamefont {L.}~\bibnamefont {Haegel}}, \bibinfo {author} {\bibfnamefont {A.}~\bibnamefont {Iess}}, \bibinfo {author} {\bibfnamefont {D.}~\bibnamefont {Keitel}},
  \bibinfo {author} {\bibfnamefont {Z.}~\bibnamefont {Márka}}, \bibinfo {author} {\bibfnamefont {S.}~\bibnamefont {Márka}}, \bibinfo {author} {\bibfnamefont {F.}~\bibnamefont {Morawski}}, \bibinfo {author} {\bibfnamefont {T.}~\bibnamefont {Nguyen}}, \bibinfo {author} {\bibfnamefont {R.}~\bibnamefont {Ormiston}}, \bibinfo {author} {\bibfnamefont {M.}~\bibnamefont {Pürrer}}, \bibinfo {author} {\bibfnamefont {M.}~\bibnamefont {Razzano}}, \bibinfo {author} {\bibfnamefont {K.}~\bibnamefont {Staats}}, \bibinfo {author} {\bibfnamefont {G.}~\bibnamefont {Vajente}}, \ and\ \bibinfo {author} {\bibfnamefont {D.}~\bibnamefont {Williams}},\ }\href {\doibase 10.1088/2632-2153/abb93a} {\bibfield  {journal} {\bibinfo  {journal} {Machine Learning: Science and Technology}\ }\textbf {\bibinfo {volume} {2}},\ \bibinfo {pages} {011002} (\bibinfo {year} {2020})}\BibitemShut {NoStop}%
\bibitem [{\citenamefont {Benedetto}\ \emph {et~al.}(2023)\citenamefont {Benedetto}, \citenamefont {Gissi}, \citenamefont {Ciaparrone},\ and\ \citenamefont {Troiano}}]{app13179886}%
  \BibitemOpen
  \bibfield  {author} {\bibinfo {author} {\bibfnamefont {V.}~\bibnamefont {Benedetto}}, \bibinfo {author} {\bibfnamefont {F.}~\bibnamefont {Gissi}}, \bibinfo {author} {\bibfnamefont {G.}~\bibnamefont {Ciaparrone}}, \ and\ \bibinfo {author} {\bibfnamefont {L.}~\bibnamefont {Troiano}},\ }\href {\doibase 10.3390/app13179886} {\bibfield  {journal} {\bibinfo  {journal} {Applied Sciences}\ }\textbf {\bibinfo {volume} {13}} (\bibinfo {year} {2023}),\ 10.3390/app13179886}\BibitemShut {NoStop}%
\bibitem [{\citenamefont {Schäfer}\ \emph {et~al.}(2023)\citenamefont {Schäfer}, \citenamefont {Zelenka}, \citenamefont {Nitz}, \citenamefont {Wang}, \citenamefont {Wu}, \citenamefont {Guo}, \citenamefont {Cao}, \citenamefont {Ren}, \citenamefont {Nousi}, \citenamefont {Stergioulas}, \citenamefont {Iosif}, \citenamefont {Koloniari}, \citenamefont {Tefas}, \citenamefont {Passalis}, \citenamefont {Salemi}, \citenamefont {Vedovato}, \citenamefont {Klimenko}, \citenamefont {Mishra}, \citenamefont {Brügmann}, \citenamefont {Cuoco}, \citenamefont {Huerta}, \citenamefont {Messenger},\ and\ \citenamefont {Ohme}}]{Sch_fer_2023}%
  \BibitemOpen
  \bibfield  {author} {\bibinfo {author} {\bibfnamefont {M.~B.}\ \bibnamefont {Schäfer}}, \bibinfo {author} {\bibfnamefont {O.}~\bibnamefont {Zelenka}}, \bibinfo {author} {\bibfnamefont {A.~H.}\ \bibnamefont {Nitz}}, \bibinfo {author} {\bibfnamefont {H.}~\bibnamefont {Wang}}, \bibinfo {author} {\bibfnamefont {S.}~\bibnamefont {Wu}}, \bibinfo {author} {\bibfnamefont {Z.-K.}\ \bibnamefont {Guo}}, \bibinfo {author} {\bibfnamefont {Z.}~\bibnamefont {Cao}}, \bibinfo {author} {\bibfnamefont {Z.}~\bibnamefont {Ren}}, \bibinfo {author} {\bibfnamefont {P.}~\bibnamefont {Nousi}}, \bibinfo {author} {\bibfnamefont {N.}~\bibnamefont {Stergioulas}}, \bibinfo {author} {\bibfnamefont {P.}~\bibnamefont {Iosif}}, \bibinfo {author} {\bibfnamefont {A.~E.}\ \bibnamefont {Koloniari}}, \bibinfo {author} {\bibfnamefont {A.}~\bibnamefont {Tefas}}, \bibinfo {author} {\bibfnamefont {N.}~\bibnamefont {Passalis}}, \bibinfo {author} {\bibfnamefont {F.}~\bibnamefont {Salemi}}, \bibinfo {author} {\bibfnamefont {G.}~\bibnamefont
  {Vedovato}}, \bibinfo {author} {\bibfnamefont {S.}~\bibnamefont {Klimenko}}, \bibinfo {author} {\bibfnamefont {T.}~\bibnamefont {Mishra}}, \bibinfo {author} {\bibfnamefont {B.}~\bibnamefont {Brügmann}}, \bibinfo {author} {\bibfnamefont {E.}~\bibnamefont {Cuoco}}, \bibinfo {author} {\bibfnamefont {E.}~\bibnamefont {Huerta}}, \bibinfo {author} {\bibfnamefont {C.}~\bibnamefont {Messenger}}, \ and\ \bibinfo {author} {\bibfnamefont {F.}~\bibnamefont {Ohme}},\ }\href {\doibase 10.1103/physrevd.107.023021} {\bibfield  {journal} {\bibinfo  {journal} {Physical Review D}\ }\textbf {\bibinfo {volume} {107}} (\bibinfo {year} {2023}),\ 10.1103/physrevd.107.023021}\BibitemShut {NoStop}%
\bibitem [{\citenamefont {Stergioulas}(2024)}]{Stergioulas:2024}%
  \BibitemOpen
  \bibfield  {author} {\bibinfo {author} {\bibfnamefont {N.}~\bibnamefont {Stergioulas}},\ }\href@noop {} {\  (\bibinfo {year} {2024})},\ \Eprint {http://arxiv.org/abs/2401.07406} {arXiv:2401.07406 [gr-qc]} \BibitemShut {NoStop}%
\bibitem [{\citenamefont {{Badger}}\ \emph {et~al.}(2023)\citenamefont {{Badger}}, \citenamefont {{Martinovic}}, \citenamefont {{Torres-Forn{\'e}}}, \citenamefont {{Sakellariadou}},\ and\ \citenamefont {{Font}}}]{Charlie+2023}%
  \BibitemOpen
  \bibfield  {author} {\bibinfo {author} {\bibfnamefont {C.}~\bibnamefont {{Badger}}}, \bibinfo {author} {\bibfnamefont {K.}~\bibnamefont {{Martinovic}}}, \bibinfo {author} {\bibfnamefont {A.}~\bibnamefont {{Torres-Forn{\'e}}}}, \bibinfo {author} {\bibfnamefont {M.}~\bibnamefont {{Sakellariadou}}}, \ and\ \bibinfo {author} {\bibfnamefont {J.~A.}\ \bibnamefont {{Font}}},\ }\href {\doibase 10.1103/PhysRevLett.130.091401} {\bibfield  {journal} {\bibinfo  {journal} {Physical Review Letters}\ }\textbf {\bibinfo {volume} {130}},\ \bibinfo {eid} {091401} (\bibinfo {year} {2023})},\ \Eprint {http://arxiv.org/abs/2210.06194} {arXiv:2210.06194 [gr-qc]} \BibitemShut {NoStop}%
\bibitem [{\citenamefont {Srinivasan}\ \emph {et~al.}(2023)\citenamefont {Srinivasan}, \citenamefont {Lamberts}, \citenamefont {Bizouard}, \citenamefont {Bruel},\ and\ \citenamefont {Mastrogiovanni}}]{Srinivasan+23}%
  \BibitemOpen
  \bibfield  {author} {\bibinfo {author} {\bibfnamefont {R.}~\bibnamefont {Srinivasan}}, \bibinfo {author} {\bibfnamefont {A.}~\bibnamefont {Lamberts}}, \bibinfo {author} {\bibfnamefont {M.~A.}\ \bibnamefont {Bizouard}}, \bibinfo {author} {\bibfnamefont {T.}~\bibnamefont {Bruel}}, \ and\ \bibinfo {author} {\bibfnamefont {S.}~\bibnamefont {Mastrogiovanni}},\ }\href {\doibase 10.1093/mnras/stad1825} {\bibfield  {journal} {\bibinfo  {journal} {Monthly Notices of the Royal Astronomical Society}\ }\textbf {\bibinfo {volume} {524}},\ \bibinfo {pages} {60} (\bibinfo {year} {2023})},\ \Eprint {http://arxiv.org/abs/https://doi.org/10.1093/mnras/stad1825} {https://doi.org/10.1093/mnras/stad1825} \BibitemShut {NoStop}%
\bibitem [{\citenamefont {Nitz}\ \emph {et~al.}(2024)\citenamefont {Nitz}, \citenamefont {Harry}, \citenamefont {Brown}, \citenamefont {Biwer}, \citenamefont {Willis}, \citenamefont {Canton}, \citenamefont {Capano}, \citenamefont {Dent}, \citenamefont {Pekowsky}, \citenamefont {Davies}, \citenamefont {De}, \citenamefont {Cabero}, \citenamefont {Wu}, \citenamefont {Williamson}, \citenamefont {Machenschalk}, \citenamefont {Macleod}, \citenamefont {Pannarale}, \citenamefont {Kumar}, \citenamefont {Reyes}, \citenamefont {dfinstad}, \citenamefont {Kumar}, \citenamefont {Tápai}, \citenamefont {Singer}, \citenamefont {Kumar}, \citenamefont {veronica villa}, \citenamefont {maxtrevor}, \citenamefont {Gadre}, \citenamefont {Khan}, \citenamefont {Fairhurst},\ and\ \citenamefont {Tolley}}]{alex_nitz_2024_10473621}%
  \BibitemOpen
  \bibfield  {author} {\bibinfo {author} {\bibfnamefont {A.}~\bibnamefont {Nitz}}, \bibinfo {author} {\bibfnamefont {I.}~\bibnamefont {Harry}}, \bibinfo {author} {\bibfnamefont {D.}~\bibnamefont {Brown}}, \bibinfo {author} {\bibfnamefont {C.~M.}\ \bibnamefont {Biwer}}, \bibinfo {author} {\bibfnamefont {J.}~\bibnamefont {Willis}}, \bibinfo {author} {\bibfnamefont {T.~D.}\ \bibnamefont {Canton}}, \bibinfo {author} {\bibfnamefont {C.}~\bibnamefont {Capano}}, \bibinfo {author} {\bibfnamefont {T.}~\bibnamefont {Dent}}, \bibinfo {author} {\bibfnamefont {L.}~\bibnamefont {Pekowsky}}, \bibinfo {author} {\bibfnamefont {G.~S.~C.}\ \bibnamefont {Davies}}, \bibinfo {author} {\bibfnamefont {S.}~\bibnamefont {De}}, \bibinfo {author} {\bibfnamefont {M.}~\bibnamefont {Cabero}}, \bibinfo {author} {\bibfnamefont {S.}~\bibnamefont {Wu}}, \bibinfo {author} {\bibfnamefont {A.~R.}\ \bibnamefont {Williamson}}, \bibinfo {author} {\bibfnamefont {B.}~\bibnamefont {Machenschalk}}, \bibinfo {author} {\bibfnamefont {D.}~\bibnamefont
  {Macleod}}, \bibinfo {author} {\bibfnamefont {F.}~\bibnamefont {Pannarale}}, \bibinfo {author} {\bibfnamefont {P.}~\bibnamefont {Kumar}}, \bibinfo {author} {\bibfnamefont {S.}~\bibnamefont {Reyes}}, \bibinfo {author} {\bibnamefont {dfinstad}}, \bibinfo {author} {\bibfnamefont {S.}~\bibnamefont {Kumar}}, \bibinfo {author} {\bibfnamefont {M.}~\bibnamefont {Tápai}}, \bibinfo {author} {\bibfnamefont {L.}~\bibnamefont {Singer}}, \bibinfo {author} {\bibfnamefont {P.}~\bibnamefont {Kumar}}, \bibinfo {author} {\bibnamefont {veronica villa}}, \bibinfo {author} {\bibnamefont {maxtrevor}}, \bibinfo {author} {\bibfnamefont {B.~U.~V.}\ \bibnamefont {Gadre}}, \bibinfo {author} {\bibfnamefont {S.}~\bibnamefont {Khan}}, \bibinfo {author} {\bibfnamefont {S.}~\bibnamefont {Fairhurst}}, \ and\ \bibinfo {author} {\bibfnamefont {A.}~\bibnamefont {Tolley}},\ }\href {\doibase 10.5281/zenodo.10473621} {\enquote {\bibinfo {title} {gwastro/pycbc: v2.3.3 release of pycbc},}\ } (\bibinfo {year} {2024})\BibitemShut {NoStop}%
\bibitem [{\citenamefont {Abbott}\ \emph {et~al.}(2018{\natexlab{a}})\citenamefont {Abbott} \emph {et~al.}}]{KAGRA:2013rdx}%
  \BibitemOpen
  \bibfield  {author} {\bibinfo {author} {\bibfnamefont {B.~P.}\ \bibnamefont {Abbott}} \emph {et~al.} (\bibinfo {collaboration} {KAGRA, LIGO Scientific, Virgo, VIRGO}),\ }\href {\doibase 10.1007/s41114-020-00026-9} {\bibfield  {journal} {\bibinfo  {journal} {Living Rev. Rel.}\ }\textbf {\bibinfo {volume} {21}},\ \bibinfo {pages} {3} (\bibinfo {year} {2018}{\natexlab{a}})},\ \Eprint {http://arxiv.org/abs/1304.0670} {arXiv:1304.0670 [gr-qc]} \BibitemShut {NoStop}%
\bibitem [{\citenamefont {{Chen}}\ \emph {et~al.}(2001)\citenamefont {{Chen}}, \citenamefont {{Donoho}},\ and\ \citenamefont {{Saunders}}}]{Chen:2001}%
  \BibitemOpen
  \bibfield  {author} {\bibinfo {author} {\bibfnamefont {S.~S.}\ \bibnamefont {{Chen}}}, \bibinfo {author} {\bibfnamefont {D.~L.}\ \bibnamefont {{Donoho}}}, \ and\ \bibinfo {author} {\bibfnamefont {M.~A.}\ \bibnamefont {{Saunders}}},\ }\href {\doibase 10.1137/S003614450037906X} {\bibfield  {journal} {\bibinfo  {journal} {SIAM Review}\ }\textbf {\bibinfo {volume} {43}},\ \bibinfo {pages} {129} (\bibinfo {year} {2001})}\BibitemShut {NoStop}%
\bibitem [{\citenamefont {{Elad}}\ and\ \citenamefont {{Aharon}}(2006)}]{Elad:2006}%
  \BibitemOpen
  \bibfield  {author} {\bibinfo {author} {\bibfnamefont {M.}~\bibnamefont {{Elad}}}\ and\ \bibinfo {author} {\bibfnamefont {M.}~\bibnamefont {{Aharon}}},\ }\href {\doibase 10.1109/TIP.2006.881969} {\bibfield  {journal} {\bibinfo  {journal} {IEEE Transactions on Image Processing}\ }\textbf {\bibinfo {volume} {15}},\ \bibinfo {pages} {3736} (\bibinfo {year} {2006})}\BibitemShut {NoStop}%
\bibitem [{\citenamefont {Mairal}\ \emph {et~al.}(2012)\citenamefont {Mairal}, \citenamefont {Bach},\ and\ \citenamefont {Ponce}}]{Mairal:2012}%
  \BibitemOpen
  \bibfield  {author} {\bibinfo {author} {\bibfnamefont {J.}~\bibnamefont {Mairal}}, \bibinfo {author} {\bibfnamefont {F.~R.}\ \bibnamefont {Bach}}, \ and\ \bibinfo {author} {\bibfnamefont {J.}~\bibnamefont {Ponce}},\ }\href {\doibase 10.1109/TPAMI.2011.156} {\bibfield  {journal} {\bibinfo  {journal} {{IEEE} Trans. Pattern Anal. Mach. Intell.}\ }\textbf {\bibinfo {volume} {34}},\ \bibinfo {pages} {791} (\bibinfo {year} {2012})}\BibitemShut {NoStop}%
\bibitem [{\citenamefont {{Torres-Forn{\'e}}}\ \emph {et~al.}(2016)\citenamefont {{Torres-Forn{\'e}}}, \citenamefont {{Marquina}}, \citenamefont {{Font}},\ and\ \citenamefont {{Ib{\'a}{\~n}ez}}}]{Alex+2016_GWDenoising}%
  \BibitemOpen
  \bibfield  {author} {\bibinfo {author} {\bibfnamefont {A.}~\bibnamefont {{Torres-Forn{\'e}}}}, \bibinfo {author} {\bibfnamefont {A.}~\bibnamefont {{Marquina}}}, \bibinfo {author} {\bibfnamefont {J.~A.}\ \bibnamefont {{Font}}}, \ and\ \bibinfo {author} {\bibfnamefont {J.~M.}\ \bibnamefont {{Ib{\'a}{\~n}ez}}},\ }\href {\doibase 10.1103/PhysRevD.94.124040} {\bibfield  {journal} {\bibinfo  {journal} {\prd}\ }\textbf {\bibinfo {volume} {94}},\ \bibinfo {eid} {124040} (\bibinfo {year} {2016})},\ \Eprint {http://arxiv.org/abs/1612.01305} {arXiv:1612.01305 [astro-ph.IM]} \BibitemShut {NoStop}%
\bibitem [{\citenamefont {{Llorens-Monteagudo}}\ \emph {et~al.}(2019)\citenamefont {{Llorens-Monteagudo}}, \citenamefont {{Torres-Forn{\'e}}}, \citenamefont {{Font}},\ and\ \citenamefont {{Marquina}}}]{Miquel:2019}%
  \BibitemOpen
  \bibfield  {author} {\bibinfo {author} {\bibfnamefont {M.}~\bibnamefont {{Llorens-Monteagudo}}}, \bibinfo {author} {\bibfnamefont {A.}~\bibnamefont {{Torres-Forn{\'e}}}}, \bibinfo {author} {\bibfnamefont {J.~A.}\ \bibnamefont {{Font}}}, \ and\ \bibinfo {author} {\bibfnamefont {A.}~\bibnamefont {{Marquina}}},\ }\href {\doibase 10.1088/1361-6382/ab0657} {\bibfield  {journal} {\bibinfo  {journal} {Classical and Quantum Gravity}\ }\textbf {\bibinfo {volume} {36}},\ \bibinfo {eid} {075005} (\bibinfo {year} {2019})},\ \Eprint {http://arxiv.org/abs/1811.03867} {arXiv:1811.03867 [astro-ph.IM]} \BibitemShut {NoStop}%
\bibitem [{\citenamefont {{Torres-Forn{\'e}}}\ \emph {et~al.}(2020)\citenamefont {{Torres-Forn{\'e}}}, \citenamefont {{Cuoco}}, \citenamefont {{Font}},\ and\ \citenamefont {{Marquina}}}]{Alex+2020}%
  \BibitemOpen
  \bibfield  {author} {\bibinfo {author} {\bibfnamefont {A.}~\bibnamefont {{Torres-Forn{\'e}}}}, \bibinfo {author} {\bibfnamefont {E.}~\bibnamefont {{Cuoco}}}, \bibinfo {author} {\bibfnamefont {J.~A.}\ \bibnamefont {{Font}}}, \ and\ \bibinfo {author} {\bibfnamefont {A.}~\bibnamefont {{Marquina}}},\ }\href {\doibase 10.1103/PhysRevD.102.023011} {\bibfield  {journal} {\bibinfo  {journal} {Physical Review D}\ }\textbf {\bibinfo {volume} {102}},\ \bibinfo {eid} {023011} (\bibinfo {year} {2020})},\ \Eprint {http://arxiv.org/abs/2002.11668} {arXiv:2002.11668 [gr-qc]} \BibitemShut {NoStop}%
\bibitem [{\citenamefont {{Saiz-P{\'e}rez}}\ \emph {et~al.}(2022)\citenamefont {{Saiz-P{\'e}rez}}, \citenamefont {{Torres-Forn{\'e}}},\ and\ \citenamefont {{Font}}}]{Saiz-Perez2022}%
  \BibitemOpen
  \bibfield  {author} {\bibinfo {author} {\bibfnamefont {A.}~\bibnamefont {{Saiz-P{\'e}rez}}}, \bibinfo {author} {\bibfnamefont {A.}~\bibnamefont {{Torres-Forn{\'e}}}}, \ and\ \bibinfo {author} {\bibfnamefont {J.~A.}\ \bibnamefont {{Font}}},\ }\href {\doibase 10.1093/mnras/stac698} {\bibfield  {journal} {\bibinfo  {journal} {Mon. Not. R. Astron. Soc}\ }\textbf {\bibinfo {volume} {512}},\ \bibinfo {pages} {3815} (\bibinfo {year} {2022})},\ \Eprint {http://arxiv.org/abs/2110.12941} {arXiv:2110.12941 [gr-qc]} \BibitemShut {NoStop}%
\bibitem [{\citenamefont {Powell}\ \emph {et~al.}(2023)\citenamefont {Powell}, \citenamefont {Iess}, \citenamefont {Llorens-Monteagudo}, \citenamefont {Obergaulinger}, \citenamefont {M\"uller}, \citenamefont {Torres-Forn\'e}, \citenamefont {Cuoco},\ and\ \citenamefont {Font}}]{Powell2023}%
  \BibitemOpen
  \bibfield  {author} {\bibinfo {author} {\bibfnamefont {J.}~\bibnamefont {Powell}}, \bibinfo {author} {\bibfnamefont {A.}~\bibnamefont {Iess}}, \bibinfo {author} {\bibfnamefont {M.}~\bibnamefont {Llorens-Monteagudo}}, \bibinfo {author} {\bibfnamefont {M.}~\bibnamefont {Obergaulinger}}, \bibinfo {author} {\bibfnamefont {B.}~\bibnamefont {M\"uller}}, \bibinfo {author} {\bibfnamefont {A.}~\bibnamefont {Torres-Forn\'e}}, \bibinfo {author} {\bibfnamefont {E.}~\bibnamefont {Cuoco}}, \ and\ \bibinfo {author} {\bibfnamefont {J.~A.}\ \bibnamefont {Font}},\ }\href@noop {} {\  (\bibinfo {year} {2023})},\ \Eprint {http://arxiv.org/abs/2311.18221} {arXiv:2311.18221 [astro-ph.HE]} \BibitemShut {NoStop}%
\bibitem [{\citenamefont {{Mallat}}\ and\ \citenamefont {{Zhang}}(1993)}]{Mallat:1993}%
  \BibitemOpen
  \bibfield  {author} {\bibinfo {author} {\bibfnamefont {S.~G.}\ \bibnamefont {{Mallat}}}\ and\ \bibinfo {author} {\bibfnamefont {Z.}~\bibnamefont {{Zhang}}},\ }\href {\doibase 10.1109/78.258082} {\bibfield  {journal} {\bibinfo  {journal} {IEEE Transactions on Signal Processing}\ }\textbf {\bibinfo {volume} {41}},\ \bibinfo {pages} {3397} (\bibinfo {year} {1993})}\BibitemShut {NoStop}%
\bibitem [{\citenamefont {Sadeghi}\ \emph {et~al.}(2013)\citenamefont {Sadeghi}, \citenamefont {Babaie-Zadeh},\ and\ \citenamefont {Jutten}}]{dict_learning}%
  \BibitemOpen
  \bibfield  {author} {\bibinfo {author} {\bibfnamefont {M.}~\bibnamefont {Sadeghi}}, \bibinfo {author} {\bibfnamefont {M.}~\bibnamefont {Babaie-Zadeh}}, \ and\ \bibinfo {author} {\bibfnamefont {C.}~\bibnamefont {Jutten}},\ }\href {\doibase 10.1109/LSP.2013.2285218} {\bibfield  {journal} {\bibinfo  {journal} {IEEE Signal Processing Letters}\ }\textbf {\bibinfo {volume} {20}},\ \bibinfo {pages} {1195} (\bibinfo {year} {2013})}\BibitemShut {NoStop}%
\bibitem [{\citenamefont {Aharon}\ and\ \citenamefont {Elad}(2008)}]{10.1137/07070156X}%
  \BibitemOpen
  \bibfield  {author} {\bibinfo {author} {\bibfnamefont {M.}~\bibnamefont {Aharon}}\ and\ \bibinfo {author} {\bibfnamefont {M.}~\bibnamefont {Elad}},\ }\href {\doibase 10.1137/07070156X} {\bibfield  {journal} {\bibinfo  {journal} {SIAM J. Img. Sci.}\ }\textbf {\bibinfo {volume} {1}},\ \bibinfo {pages} {228–247} (\bibinfo {year} {2008})}\BibitemShut {NoStop}%
\bibitem [{\citenamefont {Mairal}\ \emph {et~al.}(2009)\citenamefont {Mairal}, \citenamefont {Bach}, \citenamefont {Ponce},\ and\ \citenamefont {Sapiro}}]{Mairal:2009}%
  \BibitemOpen
  \bibfield  {author} {\bibinfo {author} {\bibfnamefont {J.}~\bibnamefont {Mairal}}, \bibinfo {author} {\bibfnamefont {F.}~\bibnamefont {Bach}}, \bibinfo {author} {\bibfnamefont {J.}~\bibnamefont {Ponce}}, \ and\ \bibinfo {author} {\bibfnamefont {G.}~\bibnamefont {Sapiro}},\ }in\ \href {\doibase 10.1145/1553374.1553463} {\emph {\bibinfo {booktitle} {Proceedings of the 26th Annual International Conference on Machine Learning}}},\ \bibinfo {series and number} {ICML '09}\ (\bibinfo  {publisher} {Association for Computing Machinery},\ \bibinfo {address} {New York, NY, USA},\ \bibinfo {year} {2009})\ p.\ \bibinfo {pages} {689–696}\BibitemShut {NoStop}%
\bibitem [{\citenamefont {Chen}\ \emph {et~al.}(2001{\natexlab{a}})\citenamefont {Chen}, \citenamefont {Donoho},\ and\ \citenamefont {Saunders}}]{BasisPursuit_Chen+2001}%
  \BibitemOpen
  \bibfield  {author} {\bibinfo {author} {\bibfnamefont {S.~S.}\ \bibnamefont {Chen}}, \bibinfo {author} {\bibfnamefont {D.~L.}\ \bibnamefont {Donoho}}, \ and\ \bibinfo {author} {\bibfnamefont {M.~A.}\ \bibnamefont {Saunders}},\ }\href {\doibase 10.1137/S003614450037906X} {\bibfield  {journal} {\bibinfo  {journal} {SIAM Review}\ }\textbf {\bibinfo {volume} {43}},\ \bibinfo {pages} {129} (\bibinfo {year} {2001}{\natexlab{a}})},\ \Eprint {http://arxiv.org/abs/https://doi.org/10.1137/S003614450037906X} {https://doi.org/10.1137/S003614450037906X} \BibitemShut {NoStop}%
\bibitem [{\citenamefont {Chen}\ \emph {et~al.}(2001{\natexlab{b}})\citenamefont {Chen}, \citenamefont {Donoho},\ and\ \citenamefont {Saunders}}]{basis_pursuit}%
  \BibitemOpen
  \bibfield  {author} {\bibinfo {author} {\bibfnamefont {S.~S.}\ \bibnamefont {Chen}}, \bibinfo {author} {\bibfnamefont {D.~L.}\ \bibnamefont {Donoho}}, \ and\ \bibinfo {author} {\bibfnamefont {M.~A.}\ \bibnamefont {Saunders}},\ }\href {http://www.jstor.org/stable/3649687} {\bibfield  {journal} {\bibinfo  {journal} {SIAM Review}\ }\textbf {\bibinfo {volume} {43}},\ \bibinfo {pages} {129} (\bibinfo {year} {2001}{\natexlab{b}})}\BibitemShut {NoStop}%
\bibitem [{\citenamefont {Tibshirani}(1996)}]{Lasso_Tibshirani1996}%
  \BibitemOpen
  \bibfield  {author} {\bibinfo {author} {\bibfnamefont {R.}~\bibnamefont {Tibshirani}},\ }\href {http://www.jstor.org/stable/2346178} {\bibfield  {journal} {\bibinfo  {journal} {Journal of the Royal Statistical Society. Series B (Methodological)}\ }\textbf {\bibinfo {volume} {58}},\ \bibinfo {pages} {267} (\bibinfo {year} {1996})}\BibitemShut {NoStop}%
\bibitem [{\citenamefont {Cutler}\ and\ \citenamefont {Flanagan}(1994)}]{Cutler:1994ys}%
  \BibitemOpen
  \bibfield  {author} {\bibinfo {author} {\bibfnamefont {C.}~\bibnamefont {Cutler}}\ and\ \bibinfo {author} {\bibfnamefont {E.~E.}\ \bibnamefont {Flanagan}},\ }\href {\doibase 10.1103/PhysRevD.49.2658} {\bibfield  {journal} {\bibinfo  {journal} {Phys. Rev. D}\ }\textbf {\bibinfo {volume} {49}},\ \bibinfo {pages} {2658} (\bibinfo {year} {1994})},\ \Eprint {http://arxiv.org/abs/gr-qc/9402014} {arXiv:gr-qc/9402014} \BibitemShut {NoStop}%
\bibitem [{\citenamefont {Cornish}\ and\ \citenamefont {Littenberg}(2015)}]{Cornish:2014kda}%
  \BibitemOpen
  \bibfield  {author} {\bibinfo {author} {\bibfnamefont {N.~J.}\ \bibnamefont {Cornish}}\ and\ \bibinfo {author} {\bibfnamefont {T.~B.}\ \bibnamefont {Littenberg}},\ }\href {\doibase 10.1088/0264-9381/32/13/135012} {\bibfield  {journal} {\bibinfo  {journal} {Class. Quant. Grav.}\ }\textbf {\bibinfo {volume} {32}},\ \bibinfo {pages} {135012} (\bibinfo {year} {2015})},\ \Eprint {http://arxiv.org/abs/1410.3835} {arXiv:1410.3835 [gr-qc]} \BibitemShut {NoStop}%
\bibitem [{\citenamefont {Cornish}\ \emph {et~al.}(2021)\citenamefont {Cornish}, \citenamefont {Littenberg}, \citenamefont {B\'ecsy}, \citenamefont {Chatziioannou}, \citenamefont {Clark}, \citenamefont {Ghonge},\ and\ \citenamefont {Millhouse}}]{Cornish:2020dwh}%
  \BibitemOpen
  \bibfield  {author} {\bibinfo {author} {\bibfnamefont {N.~J.}\ \bibnamefont {Cornish}}, \bibinfo {author} {\bibfnamefont {T.~B.}\ \bibnamefont {Littenberg}}, \bibinfo {author} {\bibfnamefont {B.}~\bibnamefont {B\'ecsy}}, \bibinfo {author} {\bibfnamefont {K.}~\bibnamefont {Chatziioannou}}, \bibinfo {author} {\bibfnamefont {J.~A.}\ \bibnamefont {Clark}}, \bibinfo {author} {\bibfnamefont {S.}~\bibnamefont {Ghonge}}, \ and\ \bibinfo {author} {\bibfnamefont {M.}~\bibnamefont {Millhouse}},\ }\href {\doibase 10.1103/PhysRevD.103.044006} {\bibfield  {journal} {\bibinfo  {journal} {Phys. Rev. D}\ }\textbf {\bibinfo {volume} {103}},\ \bibinfo {pages} {044006} (\bibinfo {year} {2021})},\ \Eprint {http://arxiv.org/abs/2011.09494} {arXiv:2011.09494 [gr-qc]} \BibitemShut {NoStop}%
\bibitem [{\citenamefont {Breivik}\ \emph {et~al.}(2020)\citenamefont {Breivik} \emph {et~al.}}]{Breivik:2019lmt}%
  \BibitemOpen
  \bibfield  {author} {\bibinfo {author} {\bibfnamefont {K.}~\bibnamefont {Breivik}} \emph {et~al.},\ }\href@noop {} {\bibfield  {journal} {\bibinfo  {journal} {Astrophys. J.}\ }\textbf {\bibinfo {volume} {898}},\ \bibinfo {pages} {71} (\bibinfo {year} {2020})},\ \Eprint {http://arxiv.org/abs/1911.00903} {arXiv:1911.00903 [astro-ph.HE]} \BibitemShut {NoStop}%
\bibitem [{\citenamefont {Owen}\ and\ \citenamefont {Sathyaprakash}(1999)}]{PhysRevD.60.022002}%
  \BibitemOpen
  \bibfield  {author} {\bibinfo {author} {\bibfnamefont {B.~J.}\ \bibnamefont {Owen}}\ and\ \bibinfo {author} {\bibfnamefont {B.~S.}\ \bibnamefont {Sathyaprakash}},\ }\href {\doibase 10.1103/PhysRevD.60.022002} {\bibfield  {journal} {\bibinfo  {journal} {Phys. Rev. D}\ }\textbf {\bibinfo {volume} {60}},\ \bibinfo {pages} {022002} (\bibinfo {year} {1999})}\BibitemShut {NoStop}%
\bibitem [{\citenamefont {Cokelaer}(2007)}]{Cokelaer:2007kx}%
  \BibitemOpen
  \bibfield  {author} {\bibinfo {author} {\bibfnamefont {T.}~\bibnamefont {Cokelaer}},\ }\href {\doibase 10.1103/PhysRevD.76.102004} {\bibfield  {journal} {\bibinfo  {journal} {Phys. Rev. D}\ }\textbf {\bibinfo {volume} {76}},\ \bibinfo {pages} {102004} (\bibinfo {year} {2007})},\ \Eprint {http://arxiv.org/abs/0706.4437} {arXiv:0706.4437 [gr-qc]} \BibitemShut {NoStop}%
\bibitem [{\citenamefont {Babak}\ \emph {et~al.}(2013)\citenamefont {Babak}, \citenamefont {Biswas}, \citenamefont {Brady}, \citenamefont {Brown}, \citenamefont {Cannon}, \citenamefont {Capano}, \citenamefont {Clayton}, \citenamefont {Cokelaer}, \citenamefont {Creighton}, \citenamefont {Dent}, \citenamefont {Dietz}, \citenamefont {Fairhurst}, \citenamefont {Fotopoulos}, \citenamefont {Gonz\'alez}, \citenamefont {Hanna}, \citenamefont {Harry}, \citenamefont {Jones}, \citenamefont {Keppel}, \citenamefont {McKechan}, \citenamefont {Pekowsky}, \citenamefont {Privitera}, \citenamefont {Robinson}, \citenamefont {Rodriguez}, \citenamefont {Sathyaprakash}, \citenamefont {Sengupta}, \citenamefont {Vallisneri}, \citenamefont {Vaulin},\ and\ \citenamefont {Weinstein}}]{PhysRevD.87.024033}%
  \BibitemOpen
  \bibfield  {author} {\bibinfo {author} {\bibfnamefont {S.}~\bibnamefont {Babak}}, \bibinfo {author} {\bibfnamefont {R.}~\bibnamefont {Biswas}}, \bibinfo {author} {\bibfnamefont {P.~R.}\ \bibnamefont {Brady}}, \bibinfo {author} {\bibfnamefont {D.~A.}\ \bibnamefont {Brown}}, \bibinfo {author} {\bibfnamefont {K.}~\bibnamefont {Cannon}}, \bibinfo {author} {\bibfnamefont {C.~D.}\ \bibnamefont {Capano}}, \bibinfo {author} {\bibfnamefont {J.~H.}\ \bibnamefont {Clayton}}, \bibinfo {author} {\bibfnamefont {T.}~\bibnamefont {Cokelaer}}, \bibinfo {author} {\bibfnamefont {J.~D.~E.}\ \bibnamefont {Creighton}}, \bibinfo {author} {\bibfnamefont {T.}~\bibnamefont {Dent}}, \bibinfo {author} {\bibfnamefont {A.}~\bibnamefont {Dietz}}, \bibinfo {author} {\bibfnamefont {S.}~\bibnamefont {Fairhurst}}, \bibinfo {author} {\bibfnamefont {N.}~\bibnamefont {Fotopoulos}}, \bibinfo {author} {\bibfnamefont {G.}~\bibnamefont {Gonz\'alez}}, \bibinfo {author} {\bibfnamefont {C.}~\bibnamefont {Hanna}}, \bibinfo {author} {\bibfnamefont
  {I.~W.}\ \bibnamefont {Harry}}, \bibinfo {author} {\bibfnamefont {G.}~\bibnamefont {Jones}}, \bibinfo {author} {\bibfnamefont {D.}~\bibnamefont {Keppel}}, \bibinfo {author} {\bibfnamefont {D.~J.~A.}\ \bibnamefont {McKechan}}, \bibinfo {author} {\bibfnamefont {L.}~\bibnamefont {Pekowsky}}, \bibinfo {author} {\bibfnamefont {S.}~\bibnamefont {Privitera}}, \bibinfo {author} {\bibfnamefont {C.}~\bibnamefont {Robinson}}, \bibinfo {author} {\bibfnamefont {A.~C.}\ \bibnamefont {Rodriguez}}, \bibinfo {author} {\bibfnamefont {B.~S.}\ \bibnamefont {Sathyaprakash}}, \bibinfo {author} {\bibfnamefont {A.~S.}\ \bibnamefont {Sengupta}}, \bibinfo {author} {\bibfnamefont {M.}~\bibnamefont {Vallisneri}}, \bibinfo {author} {\bibfnamefont {R.}~\bibnamefont {Vaulin}}, \ and\ \bibinfo {author} {\bibfnamefont {A.~J.}\ \bibnamefont {Weinstein}},\ }\href {\doibase 10.1103/PhysRevD.87.024033} {\bibfield  {journal} {\bibinfo  {journal} {Phys. Rev. D}\ }\textbf {\bibinfo {volume} {87}},\ \bibinfo {pages} {024033} (\bibinfo {year}
  {2013})}\BibitemShut {NoStop}%
\bibitem [{\citenamefont {Capano}\ \emph {et~al.}(2014)\citenamefont {Capano}, \citenamefont {Pan},\ and\ \citenamefont {Buonanno}}]{Capano_2014}%
  \BibitemOpen
  \bibfield  {author} {\bibinfo {author} {\bibfnamefont {C.}~\bibnamefont {Capano}}, \bibinfo {author} {\bibfnamefont {Y.}~\bibnamefont {Pan}}, \ and\ \bibinfo {author} {\bibfnamefont {A.}~\bibnamefont {Buonanno}},\ }\href {\doibase 10.1103/physrevd.89.102003} {\bibfield  {journal} {\bibinfo  {journal} {Physical Review D}\ }\textbf {\bibinfo {volume} {89}} (\bibinfo {year} {2014}),\ 10.1103/physrevd.89.102003}\BibitemShut {NoStop}%
\bibitem [{\citenamefont {Glanzer}\ \emph {et~al.}(2021)\citenamefont {Glanzer}, \citenamefont {Banagari}, \citenamefont {Coughlin}, \citenamefont {Zevin}, \citenamefont {Bahaadini}, \citenamefont {Rohani}, \citenamefont {Allen}, \citenamefont {Berry}, \citenamefont {Crowston}, \citenamefont {Harandi}, \citenamefont {Jackson}, \citenamefont {Kalogera}, \citenamefont {Katsaggelos}, \citenamefont {Noroozi}, \citenamefont {Osterlund}, \citenamefont {Patane}, \citenamefont {Smith}, \citenamefont {Soni},\ and\ \citenamefont {Trouille}}]{glanzer_2021_5649212}%
  \BibitemOpen
  \bibfield  {author} {\bibinfo {author} {\bibfnamefont {J.}~\bibnamefont {Glanzer}}, \bibinfo {author} {\bibfnamefont {S.}~\bibnamefont {Banagari}}, \bibinfo {author} {\bibfnamefont {S.}~\bibnamefont {Coughlin}}, \bibinfo {author} {\bibfnamefont {M.}~\bibnamefont {Zevin}}, \bibinfo {author} {\bibfnamefont {S.}~\bibnamefont {Bahaadini}}, \bibinfo {author} {\bibfnamefont {N.}~\bibnamefont {Rohani}}, \bibinfo {author} {\bibfnamefont {S.}~\bibnamefont {Allen}}, \bibinfo {author} {\bibfnamefont {C.}~\bibnamefont {Berry}}, \bibinfo {author} {\bibfnamefont {K.}~\bibnamefont {Crowston}}, \bibinfo {author} {\bibfnamefont {M.}~\bibnamefont {Harandi}}, \bibinfo {author} {\bibfnamefont {C.}~\bibnamefont {Jackson}}, \bibinfo {author} {\bibfnamefont {V.}~\bibnamefont {Kalogera}}, \bibinfo {author} {\bibfnamefont {A.}~\bibnamefont {Katsaggelos}}, \bibinfo {author} {\bibfnamefont {V.}~\bibnamefont {Noroozi}}, \bibinfo {author} {\bibfnamefont {C.}~\bibnamefont {Osterlund}}, \bibinfo {author} {\bibfnamefont {O.}~\bibnamefont
  {Patane}}, \bibinfo {author} {\bibfnamefont {J.}~\bibnamefont {Smith}}, \bibinfo {author} {\bibfnamefont {S.}~\bibnamefont {Soni}}, \ and\ \bibinfo {author} {\bibfnamefont {L.}~\bibnamefont {Trouille}},\ }\href {\doibase 10.5281/zenodo.5649212} {\enquote {\bibinfo {title} {{Gravity Spy Machine Learning Classifications of LIGO Glitches from Observing Runs O1, O2, O3a, and O3b}},}\ } (\bibinfo {year} {2021})\BibitemShut {NoStop}%
\bibitem [{\citenamefont {Campanelli}\ \emph {et~al.}(2006)\citenamefont {Campanelli}, \citenamefont {Lousto},\ and\ \citenamefont {Zlochower}}]{PhysRevD.74.041501}%
  \BibitemOpen
  \bibfield  {author} {\bibinfo {author} {\bibfnamefont {M.}~\bibnamefont {Campanelli}}, \bibinfo {author} {\bibfnamefont {C.~O.}\ \bibnamefont {Lousto}}, \ and\ \bibinfo {author} {\bibfnamefont {Y.}~\bibnamefont {Zlochower}},\ }\href {\doibase 10.1103/PhysRevD.74.041501} {\bibfield  {journal} {\bibinfo  {journal} {Phys. Rev. D}\ }\textbf {\bibinfo {volume} {74}},\ \bibinfo {pages} {041501} (\bibinfo {year} {2006})}\BibitemShut {NoStop}%
\bibitem [{\citenamefont {Ajith}\ \emph {et~al.}(2011)\citenamefont {Ajith} \emph {et~al.}}]{Ajith:2009bn}%
  \BibitemOpen
  \bibfield  {author} {\bibinfo {author} {\bibfnamefont {P.}~\bibnamefont {Ajith}} \emph {et~al.},\ }\href {\doibase 10.1103/PhysRevLett.106.241101} {\bibfield  {journal} {\bibinfo  {journal} {Phys. Rev. Lett.}\ }\textbf {\bibinfo {volume} {106}},\ \bibinfo {pages} {241101} (\bibinfo {year} {2011})},\ \Eprint {http://arxiv.org/abs/0909.2867} {arXiv:0909.2867 [gr-qc]} \BibitemShut {NoStop}%
\bibitem [{\citenamefont {Adams}\ \emph {et~al.}(2013)\citenamefont {Adams}, \citenamefont {Kochanek}, \citenamefont {Beacom}, \citenamefont {Vagins},\ and\ \citenamefont {Stanek}}]{Adams_2013}%
  \BibitemOpen
  \bibfield  {author} {\bibinfo {author} {\bibfnamefont {S.~M.}\ \bibnamefont {Adams}}, \bibinfo {author} {\bibfnamefont {C.~S.}\ \bibnamefont {Kochanek}}, \bibinfo {author} {\bibfnamefont {J.~F.}\ \bibnamefont {Beacom}}, \bibinfo {author} {\bibfnamefont {M.~R.}\ \bibnamefont {Vagins}}, \ and\ \bibinfo {author} {\bibfnamefont {K.~Z.}\ \bibnamefont {Stanek}},\ }\href {\doibase 10.1088/0004-637x/778/2/164} {\bibfield  {journal} {\bibinfo  {journal} {The Astrophysical Journal}\ }\textbf {\bibinfo {volume} {778}},\ \bibinfo {pages} {164} (\bibinfo {year} {2013})}\BibitemShut {NoStop}%
\bibitem [{\citenamefont {Abbott}\ \emph {et~al.}(2019)\citenamefont {Abbott}, \citenamefont {Abbott}, \citenamefont {Abbott}, \citenamefont {Abraham}, \citenamefont {Acernese}, \citenamefont {Ackley}, \citenamefont {Adams}, \citenamefont {Adhikari}, \citenamefont {Adya}, \citenamefont {Affeldt} \emph {et~al.}}]{abbott2019search}%
  \BibitemOpen
  \bibfield  {author} {\bibinfo {author} {\bibfnamefont {B.}~\bibnamefont {Abbott}}, \bibinfo {author} {\bibfnamefont {R.}~\bibnamefont {Abbott}}, \bibinfo {author} {\bibfnamefont {T.}~\bibnamefont {Abbott}}, \bibinfo {author} {\bibfnamefont {S.}~\bibnamefont {Abraham}}, \bibinfo {author} {\bibfnamefont {F.}~\bibnamefont {Acernese}}, \bibinfo {author} {\bibfnamefont {K.}~\bibnamefont {Ackley}}, \bibinfo {author} {\bibfnamefont {C.}~\bibnamefont {Adams}}, \bibinfo {author} {\bibfnamefont {R.}~\bibnamefont {Adhikari}}, \bibinfo {author} {\bibfnamefont {V.}~\bibnamefont {Adya}}, \bibinfo {author} {\bibfnamefont {C.}~\bibnamefont {Affeldt}},  \emph {et~al.},\ }\href@noop {} {\bibfield  {journal} {\bibinfo  {journal} {The Astrophysical Journal}\ }\textbf {\bibinfo {volume} {874}},\ \bibinfo {pages} {163} (\bibinfo {year} {2019})}\BibitemShut {NoStop}%
\bibitem [{\citenamefont {Kibble}(1976)}]{Kibble_1976}%
  \BibitemOpen
  \bibfield  {author} {\bibinfo {author} {\bibfnamefont {T.~W.~B.}\ \bibnamefont {Kibble}},\ }\href {\doibase 10.1088/0305-4470/9/8/029} {\bibfield  {journal} {\bibinfo  {journal} {Journal of Physics A: Mathematical and General}\ }\textbf {\bibinfo {volume} {9}},\ \bibinfo {pages} {1387} (\bibinfo {year} {1976})}\BibitemShut {NoStop}%
\bibitem [{\citenamefont {Abbott}\ \emph {et~al.}(2018{\natexlab{b}})\citenamefont {Abbott}, \citenamefont {Abbott}, \citenamefont {Abbott}, \citenamefont {Acernese}, \citenamefont {Ackley}, \citenamefont {Adams}, \citenamefont {Adams}, \citenamefont {Addesso}, \citenamefont {Adhikari}, \citenamefont {Adya} \emph {et~al.}}]{abbott2018constraints}%
  \BibitemOpen
  \bibfield  {author} {\bibinfo {author} {\bibfnamefont {B.~P.}\ \bibnamefont {Abbott}}, \bibinfo {author} {\bibfnamefont {R.}~\bibnamefont {Abbott}}, \bibinfo {author} {\bibfnamefont {T.~D.}\ \bibnamefont {Abbott}}, \bibinfo {author} {\bibfnamefont {F.}~\bibnamefont {Acernese}}, \bibinfo {author} {\bibfnamefont {K.}~\bibnamefont {Ackley}}, \bibinfo {author} {\bibfnamefont {C.}~\bibnamefont {Adams}}, \bibinfo {author} {\bibfnamefont {T.}~\bibnamefont {Adams}}, \bibinfo {author} {\bibfnamefont {P.}~\bibnamefont {Addesso}}, \bibinfo {author} {\bibfnamefont {R.~X.}\ \bibnamefont {Adhikari}}, \bibinfo {author} {\bibfnamefont {V.~B.}\ \bibnamefont {Adya}},  \emph {et~al.},\ }\href@noop {} {\bibfield  {journal} {\bibinfo  {journal} {Physical Review D}\ }\textbf {\bibinfo {volume} {97}},\ \bibinfo {pages} {102002} (\bibinfo {year} {2018}{\natexlab{b}})}\BibitemShut {NoStop}%
\bibitem [{\citenamefont {Hunter}(2007)}]{matplotlib}%
  \BibitemOpen
  \bibfield  {author} {\bibinfo {author} {\bibfnamefont {J.~D.}\ \bibnamefont {Hunter}},\ }\href {\doibase 10.1109/MCSE.2007.55} {\bibfield  {journal} {\bibinfo  {journal} {Computing in Science \& Engineering}\ }\textbf {\bibinfo {volume} {9}},\ \bibinfo {pages} {90} (\bibinfo {year} {2007})}\BibitemShut {NoStop}%
\bibitem [{\citenamefont {van~der Walt}\ \emph {et~al.}(2011)\citenamefont {van~der Walt}, \citenamefont {Colbert},\ and\ \citenamefont {Varoquaux}}]{numpy}%
  \BibitemOpen
  \bibfield  {author} {\bibinfo {author} {\bibfnamefont {S.}~\bibnamefont {van~der Walt}}, \bibinfo {author} {\bibfnamefont {S.~C.}\ \bibnamefont {Colbert}}, \ and\ \bibinfo {author} {\bibfnamefont {G.}~\bibnamefont {Varoquaux}},\ }\href {\doibase 10.1109/mcse.2011.37} {\bibfield  {journal} {\bibinfo  {journal} {Computing in Science \& Engineering}\ }\textbf {\bibinfo {volume} {13}},\ \bibinfo {pages} {22–30} (\bibinfo {year} {2011})}\BibitemShut {NoStop}%
\bibitem [{\citenamefont {Thain}\ \emph {et~al.}(2005)\citenamefont {Thain}, \citenamefont {Tannenbaum},\ and\ \citenamefont {Livny}}]{condor}%
  \BibitemOpen
  \bibfield  {author} {\bibinfo {author} {\bibfnamefont {D.}~\bibnamefont {Thain}}, \bibinfo {author} {\bibfnamefont {T.}~\bibnamefont {Tannenbaum}}, \ and\ \bibinfo {author} {\bibfnamefont {M.}~\bibnamefont {Livny}},\ }\href {\doibase https://doi.org/10.1002/cpe.938} {\bibfield  {journal} {\bibinfo  {journal} {Concurrency and Computation: Practice and Experience}\ }\textbf {\bibinfo {volume} {17}},\ \bibinfo {pages} {323} (\bibinfo {year} {2005})},\ \Eprint {http://arxiv.org/abs/https://onlinelibrary.wiley.com/doi/pdf/10.1002/cpe.938} {https://onlinelibrary.wiley.com/doi/pdf/10.1002/cpe.938} \BibitemShut {NoStop}%
\end{thebibliography}%

\end{document}